\documentclass[11pt]{article}
\usepackage[margin=1in]{geometry}

\usepackage[utf8]{inputenc}
\usepackage[T1]{fontenc}
\usepackage{amsmath,amssymb,amsfonts}
\usepackage{mathtools}
\usepackage{amsthm}
\usepackage{graphicx}
\usepackage{booktabs}
\usepackage{algorithm}
\usepackage{algpseudocode}
\usepackage{url}
\usepackage{cite}
\usepackage[colorlinks=true, allcolors=blue]{hyperref}
\usepackage{tikz}
\usepackage{pgfplots}
\pgfplotsset{compat=1.18}
\usetikzlibrary{positioning,shapes,shapes.symbols,arrows.meta,calc,patterns}

\definecolor{powerblue}{RGB}{31,119,180}
\definecolor{logorange}{RGB}{255,127,14}
\definecolor{validgreen}{RGB}{34,139,34}

\theoremstyle{plain}

\theoremstyle{definition}
\newtheorem{definition}{Definition}

\newcommand{\vect}[1]{\mathbf{#1}}
\DeclareMathOperator*{\argmax}{arg\,max}

\begin{document}

\title{Computational Foundations for Strategic Coopetition: Formalizing Interdependence and Complementarity}

\author{
Vik Pant\thanks{Email: vik.pant@mail.utoronto.ca} \quad Eric Yu\thanks{Email: eric.yu@utoronto.ca}\\
\\
Faculty of Information\\
University of Toronto\\
140 St George St, Toronto, ON M5S 3G6, Canada
}

\maketitle

\begin{abstract}
Coopetition refers to simultaneous cooperation and competition among actors who ``cooperate to grow the pie and compete to split it up.'' Modern socio-technical systems are characterized by strategic coopetition in which actors concomitantly cooperate to create value and compete to capture it. While conceptual modeling languages such as \textit{i*} provide rich qualitative representations of strategic dependencies, they lack mechanisms for quantitative analysis of dynamic trade-offs. Conversely, classical game theory offers mathematical rigor but strips away contextual richness. This technical report bridges this gap by developing computational foundations that formalize two critical dimensions of coopetition: interdependence and complementarity. We ground interdependence in \textit{i*} structural dependency analysis, translating depender-dependee-dependum relationships into quantitative interdependence coefficients through a structured translation framework. We formalize complementarity following Brandenburger and Nalebuff's Added Value concept, modeling synergistic value creation with validated parameterization. We integrate structural dependencies with bargaining power in value appropriation and introduce a game-theoretic formulation where Nash Equilibrium incorporates structural interdependence. Validation combines comprehensive experimental testing comprising over 22,000 trials across power and logarithmic value function specifications, demonstrating functional form robustness, with empirical application to the Samsung-Sony S-LCD joint venture (2004--2011). Under strict historical alignment scoring, logarithmic specifications achieve validation score 58/60 compared to power functions (46/60), with logarithmic specifications producing realistic cooperation increases (41\%) that align with documented S-LCD patterns while power functions produce increases (166\%) that exceed realistic bounds. Statistical significance is confirmed at $p < 0.001$ with Cohen's $d > 9$ (very large effect size). This technical report serves as the foundational reference for a coordinated research program examining strategic coopetition in requirements engineering and multi-agent systems, with companion work addressing trust dynamics, collective action, and reciprocity mechanisms.
\end{abstract}


\noindent\textbf{ArXiv Classifications:} cs.SE (Software Engineering), cs.MA (Multiagent Systems), cs.AI (Artificial Intelligence)

\section{Introduction}

The landscape of modern enterprise is increasingly defined by intricate networks of collaboration where actors simultaneously cooperate and compete. Apple and Samsung compete fiercely in the smartphone market, yet Samsung supplies critical components for Apple's iPhone. Developers on software platforms must cooperate with the platform provider to create ecosystem value while competing with other developers for user attention and revenue. Enterprise departments depend on each other for resources and capabilities while competing for budget allocations. This duality, termed strategic coopetition by Brandenburger and Nalebuff \cite{brandenburger1996co}, presents profound challenges for the analysis and design of information systems intended to support such environments.

Pant \cite{pant2021strategic} established a conceptual modeling framework for strategic coopetition, building on the \textit{i*} modeling language \cite{yu1995modelling} to capture five critical dimensions of coopetitive relationships: interdependence, complementarity, trustworthiness, reciprocity, and complex actor abstractions. This conceptual framework provided rich qualitative representations of how actors depend on each other, how they create synergistic value, and how relationships evolve over time. However, as Pant noted, conceptual models alone cannot address the quantitative strategic trade-offs that characterize real coopetitive decision-making. When should an actor invest in a partnership despite dependency risks? How much value does complementarity create, and how should it be shared? What equilibrium behaviors emerge from the interaction of cooperation incentives and competitive pressures?

Traditional conceptual modeling languages like \textit{i*} \cite{yu1995modelling} and the Goal-oriented Requirement Language (GRL) \cite{amyot2010itu} provide powerful abstractions for modeling strategic actors and their intentional relationships. The \textit{i*} framework represents actors with their goals and explicitly models dependencies through the depender-dependee-dependum triad: when Actor A (the depender) depends on Actor B (the dependee) for dependum D (a goal, task, or resource), this creates a structural relationship where A's success requires B's performance. However, while \textit{i*} offers qualitative evaluation mechanisms \cite{amyot2006towards,mussbacher2011use}, it primarily operates through hierarchical goal decomposition less suited for modeling direct, continuous influences in dynamic strategic settings.

On the other hand, classical game theory \cite{fudenberg1991game,osborne1994course} provides the mathematical rigor to analyze strategic interactions through concepts like Nash Equilibrium \cite{nash1950equilibrium} and the Shapley value \cite{shapley1953value}. Game theory excels at predicting equilibrium behaviors and computing optimal strategies. However, classical models typically assume purely self-interested actors and often omit the structural dependencies, contextual richness, and relational dynamics that characterize real business relationships. There exists a critical gap between the rich, qualitative world of conceptual modeling and the precise, quantitative world of game theory.

This technical report bridges that gap by developing computational foundations that formalize strategic coopetition in a manner grounded in both conceptual modeling principles and game-theoretic rigor. We focus on two foundational dimensions: interdependence and complementarity. These dimensions are fundamental to coopetition because interdependence captures why actors must consider each other's outcomes even while competing, and complementarity explains why collaboration can create superadditive value that purely competitive interactions cannot achieve.

Our approach develops a structured translation framework from \textit{i*} conceptual models to computational game-theoretic formalizations. This translation is non-trivial: it requires principled design decisions about how qualitative structural relationships map to quantitative mathematical functions, how to operationalize dependency criticality and goal importance, how to specify value creation functions that exhibit appropriate economic properties, and how to model value appropriation in a manner consistent with bargaining theory while remaining tractable for equilibrium analysis.

\subsection{Research Program Context}

This technical report is the foundational work in a coordinated research program on computational approaches to strategic coopetition in requirements engineering and multi-agent systems. The complete program addresses five key dimensions of coopetitive relationships identified by Pant \cite{pant2021strategic}. This report establishes the core mathematical framework by formalizing interdependence and complementarity. Complementary research in this program examines trust dynamics through computational models of reliability beliefs and their evolution through repeated interactions, analyzes collective action scenarios including free-riding problems and loyalty mechanisms in collaborative software engineering, and develops reciprocity mechanisms for sequential cooperation in multi-agent coopetition. These companion studies build upon the foundational concepts established here and are forthcoming as technical reports on arXiv under the same research program.

This technical report presents an integrated computational formalization of the conceptual framework for strategic coopetition established in Pant's doctoral dissertation \cite{pant2021strategic}. While that prior work identified interdependence and complementarity as critical dimensions of coopetition and conceptually outlined how \textit{i*} modeling could articulate interdependence and how value modeling could express complementarity, the present report contributes the specific mathematical equations (Equations 1--15), the operational translation framework with explicit parameter elicitation guidance (Section 5), and the rigorous dual-track validation methodology (Sections 7--8) necessary for quantitative analysis and empirical application. The conceptual foundations are established in \cite{pant2021strategic}; this report provides their computational realization.

\subsection{Contributions}

The main contributions of this technical report are:

\begin{enumerate}
    \item A specific utility function (Equation 13) that computationally integrates the interdependence and complementarity dimensions identified in \cite{pant2021strategic}, enabling game-theoretic equilibrium analysis of coopetitive scenarios.
    
    \item A formal mathematical specification (Equation 1) and validation of the interdependence formalization approach conceptually outlined in \cite{pant2021strategic}, providing explicit translation from \textit{i*} depender-dependee-dependum relationships to quantitative interdependence coefficients.
    
    \item A structured translation framework providing operational guidance for requirements engineers and system analysts to move from qualitative \textit{i*} models to quantitative game-theoretic models, emphasizing the iterative and reflexive nature of this modeling process.
    
    \item Formalization of complementarity following Brandenburger and Nalebuff's Added Value concept, with systematically validated value creation functions demonstrating that multiple functional forms (power, logarithmic) capture coopetitive dynamics, with empirical validation identifying optimal specifications for specific contexts (logarithmic $\theta=20$ for S-LCD case).
    
    \item Integration of structural dependencies with bargaining power in value appropriation, connecting \textit{i*} dependencies to Shapley-inspired value allocation.
    
    \item A game-theoretic formulation where Nash Equilibrium incorporates structural interdependence through dependency-augmented utility functions, extending classical equilibrium concepts to account for instrumental organizational coupling.
    
    \item Comprehensive dual-track validation methodology combining experimental robustness testing across functional specifications (demonstrating framework predictions hold regardless of form) with empirical case study application (demonstrating framework captures real coopetitive dynamics in Samsung-Sony S-LCD joint venture).
\end{enumerate}

This technical report establishes the foundation for the broader research program. By focusing on interdependence and complementarity in this foundational work, we establish core concepts and validation methodology that related research builds upon.

\section{Background and Related Work}

\textit{Note: Portions of this background review adapt material from \cite{pant2021strategic} with the author's permission, providing context for the computational formalization developed in subsequent sections.}

\subsection{Conceptual Modeling of Strategic Actors}

The \textit{i*} framework, developed by Eric Yu \cite{yu1995modelling}, provides a visual modeling language for representing strategic actors, their intentional goals, and dependencies between actors. In \textit{i*}, an actor is modeled as an intentional entity with goals it seeks to achieve. The framework captures strategic dependencies through a triadic relationship: a \emph{depender} (the actor who depends), a \emph{dependee} (the actor who is depended upon), and a \emph{dependum} (the goal, task, or resource that is the object of the dependency).

This dependency structure captures instrumental interdependence in a fundamentally different way than preference-based models. When Actor A depends on Actor B for resource R, this represents a structural constraint: A cannot achieve certain goals without B successfully providing R. This is distinct from A having social preferences or altruistic concern for B's welfare. The dependency is rooted in the causal structure of goal achievement, not psychological disposition.

The \textit{i*} framework has been extensively developed for requirements engineering \cite{yu2011social}, with extensions including quantitative evaluation mechanisms. The Goal-oriented Requirement Language (GRL) \cite{amyot2010itu}, standardized by ITU-T, builds on \textit{i*} concepts and provides evaluation algorithms for propagating satisfaction levels through goal models. Researchers have developed quantitative extensions \cite{amyot2006towards,franch2011systematic,horkoff2012interactive} that assign numerical weights to goals and compute satisfaction scores. However, these approaches typically rely on hierarchical propagation through AND/OR decomposition trees and are less suited for modeling direct, continuous strategic interactions where actors simultaneously optimize their actions.

For our purposes, \textit{i*} provides the conceptual foundation for identifying and structuring dependencies, but we require translation to a continuous optimization framework for equilibrium analysis. The challenge lies in systematically extracting quantitative parameters from qualitative \textit{i*} models in a manner that preserves semantic intent while enabling mathematical analysis.

\subsection{Game Theory and Strategic Interaction}

Game theory, founded by Von Neumann and Morgenstern \cite{von1944theory}, provides the mathematical framework for analyzing strategic interactions among rational decision-makers. The central solution concept is Nash Equilibrium \cite{nash1950equilibrium,nash1951non}: an action profile where no actor can improve their payoff by unilaterally deviating. Nash Equilibrium has become the dominant paradigm for predicting strategic behavior in economics, computer science, and multi-agent systems \cite{shoham2008multiagent}.

Cooperative game theory addresses value distribution in coalitions. The Shapley value \cite{shapley1953value}, one of the most celebrated concepts in game theory, provides a principled allocation of coalition value based on each player's marginal contribution across all possible coalitions. The Shapley value satisfies desirable properties including efficiency (allocating all value), symmetry (identical players receive equal shares), null player (players contributing nothing receive nothing), and additivity. Myerson \cite{myerson1991game} and Roth \cite{roth1988shapley} have extensively developed the theory and applications of cooperative solution concepts.

However, classical game theory faces limitations when applied to real coopetitive scenarios. First, the standard assumption of purely self-interested payoffs fails to capture how structural dependencies create genuine concern for partner outcomes. Second, game-theoretic models typically specify payoff functions exogenously without connecting them to underlying organizational or technological structures that conceptual models represent. Third, the distinction between value creation and value appropriation, which is central to coopetition \cite{brandenburger1996co}, is often not explicitly modeled in classical formulations.

Our approach addresses these limitations by augmenting game-theoretic utilities with terms derived from structural dependencies, explicitly modeling value creation separately from appropriation, and grounding parameter values in conceptual modeling analysis.

\subsection{Coopetition and Value Creation}

Brandenburger and Nalebuff's seminal work \cite{brandenburger1996co} popularized the term "co-opetition" as a description of relationships exhibiting both cooperative and competitive elements. Their framework distinguishes two distinct processes: cooperating to grow the value pie (value creation) and competing to split the pie (value appropriation). This distinction is crucial for understanding coopetitive dynamics.

Complementarity plays a central role in coopetition. When actors possess heterogeneous resources or capabilities, combining them can create superadditive value: the whole exceeds the sum of the parts. Formally, complementarity exists when $V(\{i,j\}) > V(\{i\}) + V(\{j\})$, where $V(\cdot)$ measures the value created by a coalition. This superadditivity creates the cooperative incentive in coopetition because actors have reason to collaborate when they can jointly create more value than they could independently.

The coopetition literature has grown substantially. Bengtsson and Kock \cite{bengtsson2000coopetition} studied coopetition in business networks, identifying how firms cooperate on activities far from customers while competing on activities close to customers. Gnyawali and Park \cite{gnyawali2011co} examined coopetition between technology giants, analyzing the tensions and balancing mechanisms. Lado, Boyd, and Hanlon \cite{lado1997competition} developed a framework distinguishing competition and cooperation as orthogonal dimensions rather than endpoints of a continuum.

Despite this rich conceptual development, the coopetition literature has remained largely qualitative. Quantitative models of how complementarity affects strategic behavior, how value creation and appropriation interact in equilibrium, and how structural dependencies influence coopetitive outcomes are limited. Our work formalizes these dynamics in a computational framework suitable for analysis and prediction.

\subsection{Positioning This Work}

This technical report synthesizes insights from conceptual modeling, game theory, and coopetition research. From \textit{i*}, we adopt the structural dependency framework and its representation of instrumental interdependence. From game theory, we adopt equilibrium analysis and optimization-based solution concepts. From coopetition theory, we adopt the value creation versus appropriation distinction and the concept of complementarity. The synthesis produces a computational framework that maintains the semantic richness of conceptual models while enabling the quantitative analysis of game theory.

\section{Foundational Concepts}

Before presenting our mathematical formalization, we establish clear definitions of the two dimensions this technical report addresses: interdependence and complementarity.

\begin{definition}[Interdependence]
Interdependence captures the structural coupling of actors' outcomes through dependency relationships. In the \textit{i*} framework, actor $i$ (depender) depends on actor $j$ (dependee) for dependum $d$ (a goal, task, or resource). This creates instrumental interdependence: actor $i$'s goal achievement structurally requires actor $j$'s successful performance in delivering $d$. The strength of interdependence reflects the criticality and importance of these dependencies.
\end{definition}

This definition builds on the conceptual framework established in \cite{pant2021strategic}, which identified interdependence as a primary characteristic for modeling coopetition through \textit{i*} strategic actor modeling. Our contribution is the mathematical specification that operationalizes this concept for quantitative analysis.

It is crucial to distinguish instrumental interdependence from social preferences or psychological altruism. When a software development firm depends on a platform provider for API access, this dependency is structural, meaning that certain development goals literally cannot be achieved without the platform's cooperation. This differs fundamentally from the firm having prosocial preferences or caring about the platform provider's welfare for ethical reasons. Instrumental interdependence creates rational incentives to consider partner outcomes because those outcomes directly affect one's own goal achievement through causal mechanisms, not through utility function parameters representing tastes for others' wellbeing.

In \textit{i*}, dependencies are asymmetric: Actor A depending on Actor B does not imply the reverse. A mobile app developer may critically depend on the smartphone operating system provider, while the OS provider's dependence on any single app developer is negligible. This asymmetry has profound strategic implications, as it affects bargaining power and vulnerability to opportunistic behavior.

\begin{definition}[Complementarity]
Complementarity refers to the superadditive value created when actors combine distinct resources or capabilities. Following Brandenburger and Nalebuff \cite{brandenburger1996co}, complementarity exists when $V(\{i,j\}) > V(\{i\}) + V(\{j\})$, where $V(\cdot)$ represents the value created by a coalition. This superadditivity means that actors working together can create more total value than the sum of what each could create independently.
\end{definition}

Following the framework in \cite{pant2021strategic}, we formalize complementarity as superadditive value creation, operationalizing the conceptual insight that coopeting actors create synergistic value exceeding the sum of independent contributions. Our contribution is developing specific value creation functions with validated parameterization.

Complementarity relates closely to the Shapley value \cite{shapley1953value} from cooperative game theory. The Shapley value measures each player's average marginal contribution across all possible coalition formation orders. When actors are highly complementary, their Shapley values typically exceed what they could achieve alone, creating incentive to collaborate. However, complementarity alone does not determine value distribution since bargaining power, dependency structures, and contractual arrangements also matter.

Consider a platform ecosystem where the platform provider creates infrastructure value independently, and app developers create application value independently, but the combination creates network effects and user value that exceeds the sum of independent contributions. This complementarity drives the cooperative aspect of platform coopetition, while the competition for revenue share drives the competitive aspect.

\section{Mathematical Formalization}

We now develop the formal mathematical model integrating interdependence and complementarity. This section constitutes the technical core of our contribution, showing how to translate from conceptual models to computational representations.

\subsection{Notation and Basic Setup}

Consider a system of $N$ actors, indexed by $i \in \{1, \ldots, N\}$. Each actor $i$ chooses an action $a_i \in A_i$ from their action set. We focus on continuous action spaces representing investment levels or resource allocations, where $a_i \in \mathbb{R}_+$ denotes the amount of resources actor $i$ commits to the coopetitive endeavor. An action profile $\vect{a} = (a_1, \ldots, a_N) \in A_1 \times \cdots \times A_N$ represents all actors' actions simultaneously. We denote $\vect{a}_{-i}$ as the action profile excluding actor $i$, representing what all other actors do.

Following the coopetition framework \cite{brandenburger1996co}, we distinguish value creation from value appropriation. The \textbf{value creation function} $V(\vect{a})$ represents the total value generated by actors' joint actions before any distribution occurs. This total value depends on everyone's actions and captures both individual contributions and synergistic effects from collaboration.

\subsection{Formalizing Interdependence through \textit{i*} Structural Dependencies}

We formalize interdependence based on the \textit{i*} conceptual modeling framework \cite{yu1995modelling}, which captures strategic dependencies between actors through the depender-dependee-dependum relationship. The approach of using \textit{i*} to articulate interdependence was conceptually outlined in \cite{pant2021strategic}; we provide here its formal mathematical specification through Equation 1 and the structured translation methodology detailed in Section 5.

\subsubsection{The Interdependence Matrix}

The \textbf{Interdependence Matrix} $D$ is an $N \times N$ matrix where element $D_{ij}$ represents the structural dependency of actor $i$ on actor $j$. This coefficient quantifies how much actor $i$'s outcome depends on actor $j$'s actions. The matrix is generally asymmetric: $D_{ij} \neq D_{ji}$ in typical coopetitive scenarios, reflecting that dependency relationships are directional.

\subsubsection{Translation from \textit{i*} Dependency Networks}

The translation from \textit{i*} models to the interdependence matrix proceeds through several components. Let $\mathcal{D}_i$ denote the set of dependums (goals, tasks, resources) that actor $i$ seeks to achieve or obtain. In an \textit{i*} model, these dependums are identified through goal analysis and requirements elicitation. For each dependum $d \in \mathcal{D}_i$, we define:

\textbf{Importance Weight} $w_d \geq 0$: This quantifies the strategic importance or priority of dependum $d$ to actor $i$. High-importance goals receive larger weights, reflecting their criticality to the actor's overall objectives. These weights can be elicited through techniques from multi-criteria decision analysis, such as the Analytic Hierarchy Process (AHP) \cite{saaty1980analytic}, where stakeholders provide pairwise comparisons of goal importance, or through direct assessment methods where experts assign numerical priorities.

\textbf{Dependency Indicator} $\text{Dep}(i,j,d) \in \{0,1\}$: This binary indicator equals 1 if actor $i$ depends on actor $j$ for dependum $d$ (there exists a dependency link in the \textit{i*} model), and 0 otherwise. This directly reflects the \textit{i*} dependency structure.

\textbf{Criticality Factor} $\text{crit}(i,j,d) \in [0,1]$: This quantifies how critical actor $j$ is for actor $i$ achieving dependum $d$. The criticality depends on whether alternatives exist:

\begin{itemize}
    \item $\text{crit}(i,j,d) = 1$ if actor $j$ is the sole provider of $d$ with no alternatives (complete criticality)
    \item $\text{crit}(i,j,d) = 1/n$ if $n$ alternative providers exist for $d$ (criticality diminishes with alternatives)
    \item $\text{crit}(i,j,d) = \rho_d \in [0,1]$ if $d$ is partially substitutable, where $\rho_d$ reflects the degree of substitutability (lower values indicate easier substitution)
\end{itemize}

The criticality factor captures a fundamental aspect of dependency vulnerability: an actor with monopoly control over a critical resource has high criticality, while an actor providing an easily substitutable resource has low criticality even if there is a dependency relationship.

The structural interdependence coefficient is then computed as:

\begin{equation}
\label{eq:interdependence}
D_{ij} = \frac{\sum_{d \in \mathcal{D}_i} w_d \cdot \text{Dep}(i,j,d) \cdot \text{crit}(i,j,d)}{\sum_{d \in \mathcal{D}_i} w_d}
\end{equation}

\subsubsection{Interpretation and Properties}

The interdependence coefficient $D_{ij} \in [0,1]$ is normalized by the sum of importance weights, ensuring it represents a proportion of actor $i$'s total goal importance that depends on actor $j$. Key interpretations:

\begin{itemize}
    \item $D_{ij} = 0$: Actor $i$ has no dependencies on actor $j$, or all dependencies are for unimportant goals, or all dependencies have readily available alternatives.
    
    \item $D_{ij} = 1$: All of actor $i$'s important goals critically depend on actor $j$ with no alternatives. This represents complete dependency and maximum vulnerability.
    
    \item $0 < D_{ij} < 1$: Partial dependency, the typical case. The magnitude reflects the proportion of $i$'s important goals that depend critically on $j$.
    
    \item $D_{ij} \neq D_{ji}$ in general: Dependencies are asymmetric. A small startup might depend heavily on a platform provider ($D_{\text{startup},\text{platform}}$ large), while the platform's dependence on any single startup is negligible ($D_{\text{platform},\text{startup}}$ near zero).
    
    \item By convention, $D_{ii} = 0$: Actors do not depend on themselves in this formulation, though self-dependencies could be modeled through internal goal decomposition if needed.
\end{itemize}

This formalization preserves the semantic richness of \textit{i*} dependency analysis while producing quantitative coefficients suitable for utility function parameterization. The formula aggregates multiple dependencies weighted by importance and moderated by criticality, producing a single coefficient that captures the overall structural coupling between two actors.

\subsection{Formalizing Complementarity as Value Creation}

Complementarity is modeled as an intrinsic property of the value creation function. Building on the conceptual foundation in \cite{pant2021strategic} that identified value modeling for expressing complementarity, we develop specific value creation functions with validated parameterization. Following Brandenburger and Nalebuff's Added Value concept \cite{brandenburger1996co}, we seek a value function that exhibits superadditivity: joint action creates more value than the sum of individual contributions.

\subsubsection{Value Function Specification}

We parameterize the value creation function to control the degree of complementarity using parameter $\gamma \geq 0$:

\begin{equation}
\label{eq:value_function}
V(\vect{a} \mid \gamma) = \sum_{i=1}^{N} f_i(a_i) + \gamma \cdot g(a_1, \ldots, a_N)
\end{equation}

\textbf{Components:}

\begin{itemize}
    \item $f_i(a_i)$: Individual value contribution function for actor $i$, representing value that actor $i$ creates independently through their own action. This captures value that would exist even without collaboration.
    
    \item $g(a_1, \ldots, a_N)$: Synergy function, representing value that exists only through the interaction of multiple actors' actions. This is the essence of complementarity.
    
    \item $\gamma \geq 0$: Complementarity parameter controlling the degree of superadditivity. When $\gamma = 0$, value is purely additive with no complementarity. As $\gamma$ increases, synergistic effects become more significant.
\end{itemize}

\subsubsection{Functional Form Justification: Power Functions}

The choice of functional forms for $f_i$ and $g$ should reflect economic properties and domain characteristics. For investment scenarios with diminishing returns to individual effort, we instantiate:

\begin{equation}
\label{eq:individual_contribution}
f_i(a_i) = a_i^\beta \quad \text{where } \beta \in (0,1)
\end{equation}

The parameter $\beta < 1$ ensures diminishing marginal returns: the first unit of investment creates more value than the tenth unit. This power function form aligns with Cobb-Douglas production functions widely used in economics \cite{douglas1928theory}, where $\beta$ represents output elasticity with respect to input. Our validation (Section \ref{sec:validation}) demonstrates that $\beta = 0.75$ achieves optimal balance across multiple criteria for the power function specification.

For the synergy function, we require that all actors must contribute for synergy to exist (if any $a_i = 0$, then $g = 0$), and that synergy increases with each actor's contribution. The geometric mean satisfies these properties:

\begin{equation}
\label{eq:synergy_function}
g(a_1, \ldots, a_N) = (a_1 \cdot a_2 \cdot \ldots \cdot a_N)^{1/N}
\end{equation}

For the two-actor case commonly used in experiments, this simplifies to:

\begin{equation}
\label{eq:synergy_two_actor}
g(a_1, a_2) = \sqrt{a_1 \cdot a_2}
\end{equation}

The geometric mean ensures that synergy is symmetric (order doesn't matter) and requires balanced contributions (extreme imbalance reduces synergy). This captures the intuition that complementarity requires genuine collaboration, and that one actor contributing heavily while another free-rides produces less synergy than balanced contributions.

\subsubsection{Alternative Value Function Specifications}
\label{sec:logarithmic_spec}

While the power function provides theoretical tractability and aligns with Cobb-Douglas production traditions, empirical validation reveals that alternative functional forms may achieve better-performing fit with observed coopetitive behaviors depending on context-specific value creation patterns. We introduce the logarithmic value function as an empirically validated alternative specification.

For scenarios where diminishing returns manifest differently than power functions capture, the logarithmic individual contribution function offers an alternative:

\begin{equation}
\label{eq:individual_log}
f_i(a_i) = \theta \cdot \ln(1 + a_i) \quad \text{where } \theta > 0
\end{equation}

The parameter $\theta$ controls the rate of value growth. The logarithmic form exhibits different diminishing returns properties compared to power functions: initial returns diminish more rapidly but never reach zero even for very large investments. This can better represent contexts where baseline capabilities are highly valuable but incremental improvements have declining impact.

The synergy function remains unchanged as the geometric mean, maintaining the requirement for balanced contributions. The complete logarithmic value function becomes:

\begin{equation}
\label{eq:value_log}
V(\vect{a} \mid \gamma, \theta) = \sum_{i=1}^{N} \theta \cdot \ln(1 + a_i) + \gamma \cdot g(a_1, \ldots, a_N)
\end{equation}

Our empirical validation (Section \ref{sec:empirical_validation}) demonstrates that for the Samsung-Sony S-LCD joint venture case, the logarithmic specification with $\theta = 20$ achieves validation score 58/60 under strict historical alignment scoring. Power function specifications with $\beta = 0.75$ achieve 46/60, demonstrating a substantial logarithmic advantage of 12 criteria. The difference stems primarily from historical alignment: the logarithmic specification produces realistic 41\% cooperation increases that fall within the documented 15-50\% range for S-LCD, whereas the power function produces 166\% increases that exceed realistic bounds. This suggests that functional form selection should be informed by the specific empirical context, with practitioners selecting specifications that produce realistic cooperation magnitudes for their domain.

\subsubsection{Superadditivity Verification}

To verify that these value functions exhibit complementarity, consider two actors choosing actions $(a_1, a_2)$. For the power function specification, the value created jointly is:

\begin{equation}
V(\{a_1, a_2\}) = a_1^\beta + a_2^\beta + \gamma \sqrt{a_1 \cdot a_2}
\end{equation}

The value each could create independently is:

\begin{equation}
V(\{a_1\}) = a_1^\beta, \quad V(\{a_2\}) = a_2^\beta
\end{equation}

Superadditivity requires $V(\{a_1, a_2\}) > V(\{a_1\}) + V(\{a_2\})$, which holds when:

\begin{equation}
\gamma \sqrt{a_1 \cdot a_2} > 0
\end{equation}

This is satisfied for any $\gamma > 0$ and positive actions, confirming that the synergy term creates Added Value beyond what actors contribute individually. The magnitude of this added value scales with $\gamma$ (degree of complementarity) and with the geometric mean of actions (extent of collaboration). The same superadditivity property holds for logarithmic specifications.

\subsection{Value Appropriation in Coopetition}

A critical aspect of coopetition is how value gets distributed. The coopetition paradox involves two distinct processes \cite{brandenburger1996co}: cooperating to grow the pie (value creation via $V$) and competing to split the pie (value appropriation). In real coopetitive relationships, actors appropriate individually-created value while negotiating shares of synergistic value based on relative bargaining power.

\subsubsection{Private Payoff Function}

We model value appropriation through the \textbf{private payoff function}:

\begin{equation}
\label{eq:payoff}
\pi_i(\vect{a}) = e_i - a_i + f_i(a_i) + \alpha_i \left[V(\vect{a}) - \sum_{j=1}^{N} f_j(a_j)\right]
\end{equation}

\textbf{Interpretation of Terms:}

\begin{enumerate}
    \item $e_i$: Actor $i$'s initial endowment or baseline payoff before the coopetitive interaction.
    
    \item $-a_i$: The cost of actor $i$'s investment. Resources committed to the coopetitive endeavor cannot be used elsewhere.
    
    \item $f_i(a_i)$: Value that actor $i$ appropriates from their individual production. We assume actors fully capture the value they create independently. This reflects that individual contributions have clear attribution.
    
    \item $\alpha_i [V(\vect{a}) - \sum_{j} f_j(a_j)]$: Actor $i$'s share of the synergistic value. The term $S(\vect{a}) = V(\vect{a}) - \sum_{j} f_j(a_j) = \gamma \cdot g(\vect{a})$ represents total synergy created through collaboration. Actor $i$ receives share $\alpha_i$ of this synergy, determined by their bargaining power.
\end{enumerate}

This payoff structure separates individual and synergistic value appropriation. Individual value flows to its creator automatically, while synergistic value must be allocated through negotiation or institutional mechanisms. This aligns with empirical observations in coopetitive relationships: actors can easily claim credit for their own contributions, but synergistic value created through collaboration requires explicit allocation agreements.

\subsubsection{Bargaining Power and Pre-Negotiated Shares}

The allocation weights $\alpha_i$ represent actor $i$'s negotiated share of synergy, determined by their structural bargaining power. Drawing on concepts from cooperative game theory \cite{shapley1953value,myerson1991game}, we model these shares as pre-negotiated based on each actor's structural position:

\begin{equation}
\label{eq:bargaining_weights}
\alpha_i = \frac{\beta_i}{\sum_{j=1}^{N} \beta_j}
\end{equation}

where $\beta_i > 0$ represents actor $i$'s structural bargaining power parameter. This normalization ensures $\sum_{i=1}^{N} \alpha_i = 1$, fully allocating all synergistic value.

The bargaining power parameter $\beta_i$ can be informed by multiple factors:

\begin{itemize}
    \item \textbf{Market position}: Actors with strong market positions (large user bases, established brands, network effects) have higher bargaining power.
    
    \item \textbf{Dependency asymmetry}: If others depend heavily on actor $i$ but $i$ has low dependence on others, then $i$ has high bargaining power. This connects bargaining power to the interdependence matrix.
    
    \item \textbf{Outside options}: Actors with strong BATNAs (Best Alternative To Negotiated Agreement) have higher bargaining power.
    
    \item \textbf{Shapley value approximation}: The Shapley value from cooperative game theory provides a principled starting point for estimating $\beta_i$, as it measures marginal contribution to coalition value.
\end{itemize}

An important modeling choice is using pre-negotiated shares rather than computing allocations dynamically. This reflects that in many real coopetitive relationships, actors establish contractual terms (revenue sharing agreements, licensing fees, transfer pricing) before taking actions, consistent with contract theory \cite{bolton2005contract}. The shares are determined ex ante based on bargaining power, and actors then optimize their actions taking these shares as given. This is appropriate for settings with explicit contracts, platform terms of service, or established business relationships. Alternative formulations computing allocations ex post as functions of realized outcomes are possible but add substantial complexity.

\subsection{The Integrated Utility Function}

Having formalized value creation (through $V$), value appropriation (through $\pi_i$), and structural dependencies (through $D$), we now integrate these components into a unified utility function.

\begin{equation}
\label{eq:utility_integrated}
U_i(\vect{a}) = \pi_i(\vect{a}) + \sum_{j \neq i} D_{ij} \cdot \pi_j(\vect{a})
\end{equation}

\textbf{Interpretation:}

The first term $\pi_i(\vect{a})$ is actor $i$'s private payoff, representing direct returns from their own investment and their share of synergistic value. This is what a purely self-interested actor would maximize.

The second term $\sum_{j \neq i} D_{ij} \cdot \pi_j(\vect{a})$ captures instrumental interdependence. Actor $i$ rationally cares about actor $j$'s payoff $\pi_j$ proportional to the structural dependency $D_{ij}$. When $D_{ij}$ is large (actor $i$ depends heavily on $j$), actor $i$ has strong incentive to ensure $j$ succeeds, because $j$'s success is instrumentally necessary for $i$'s own goal achievement. This concern for $j$'s success is not altruism but rather represents rational self-interest in the context of structural coupling.

This utility formulation extends classical game theory by incorporating dependency-based other-regarding preferences derived from interdependency structure rather than assuming exogenously given preferences. The interdependence terms create positive spillovers that can shift equilibria toward more cooperative outcomes compared to purely self-interested Nash equilibria.

\section{Translation Framework: From \textit{i*} Models to Game-Theoretic Formalizations}

A key contribution of this technical report is providing operational guidance for translating qualitative \textit{i*} conceptual models into quantitative game-theoretic representations. This section presents a structured translation framework that requirements engineers and information systems analysts can follow.

An important characteristic of this translation framework is its \textbf{iterative and reflexive nature}. While presented as sequential steps for pedagogical clarity, real-world application involves cycling between conceptual and computational modeling. The act of quantification often reveals gaps, inconsistencies, or opportunities for refinement in the qualitative \textit{i*} model. For example, attempting to assess bargaining power (Step 7) may reveal that a critical source of leverage was not captured as an explicit dependency in the initial \textit{i*} diagram (Step 1), necessitating model revision. Similarly, computational equilibrium analysis may suggest strategic alternatives not considered in the original conceptual model, prompting analysts to expand or restructure the \textit{i*} representation. This \textbf{mutual refinement between qualitative and quantitative representations} is a strength of the approach, enabling deeper understanding through multiple modeling perspectives.

This framework provides structured guidance for parameter elicitation while recognizing that expert judgment is required at key steps. Different stages of the translation process require varying levels of domain expertise and firm-specific knowledge for successful application.

\subsection{Step-by-Step Translation Process}

\textbf{Step 1: Elicit the \textit{i*} Dependency Network}

Begin with standard \textit{i*} modeling techniques \cite{yu1995modelling,yu2011social}. Through stakeholder interviews, document analysis, and requirements workshops, identify:

\begin{itemize}
    \item Actors in the system and their boundaries
    \item Goals each actor seeks to achieve
    \item Dependencies between actors: for each goal $g$ that actor $i$ pursues, determine if $i$ depends on other actors to achieve $g$, and if so, identify the dependee and the nature of the dependum
\end{itemize}

Create the standard \textit{i*} Strategic Dependency (SD) model showing actors as circles, goals as rounded rectangles, and dependency links as arrows from depender to dependee labeled with the dependum.

\textbf{Step 2: Quantify Importance Weights}

For each actor $i$, quantify the importance weight $w_d$ for each dependum $d \in \mathcal{D}_i$. Several methods are available:

\begin{itemize}
    \item \textbf{Analytic Hierarchy Process (AHP)}: Conduct pairwise comparisons where stakeholders assess the relative importance of goals. AHP produces a priority vector through eigenvalue analysis \cite{saaty1980analytic}.
    
    \item \textbf{Direct Assessment}: Ask stakeholders to allocate 100 points across their goals reflecting relative importance. Normalize to ensure consistency.
    
    \item \textbf{Goal Criticality Analysis}: Rate goals on scales for urgency (how time-sensitive), impact (consequences of failure), and stakeholder priority. Combine ratings into an importance score.
\end{itemize}

Document the rationale for importance assignments to ensure traceability and enable sensitivity analysis.

\textbf{Elicitation Note}: Importance weights are inherently subjective and context-dependent. Different stakeholder groups may assign divergent priorities to the same goals. Multi-stakeholder workshops using techniques like AHP enable structured negotiation toward consensus weights while documenting areas of disagreement that may indicate organizational tensions requiring governance attention.

\textbf{Step 3: Assess Criticality Factors}

For each dependency relationship $\text{Dep}(i,j,d) = 1$ in the \textit{i*} model, assess the criticality factor $\text{crit}(i,j,d)$:

\begin{itemize}
    \item Identify whether actor $j$ is the sole provider of dependum $d$, or whether alternatives exist. If sole provider, set $\text{crit}(i,j,d) = 1$.
    
    \item If $n$ alternative providers exist, assess whether they are perfect substitutes (set $\text{crit}(i,j,d) = 1/n$) or if actor $j$ has advantages (quality, reliability, switching costs). If $j$ has advantages, use a value between $1/n$ and $1$ reflecting their relative criticality.
    
    \item For partially substitutable resources, estimate the substitutability parameter $\rho_d \in [0,1]$, where lower values indicate easier substitution. This might be informed by switching costs, technological compatibility, or institutional constraints.
\end{itemize}

\textbf{Elicitation Note}: Criticality assessment requires market intelligence about alternative suppliers, technological substitutability, and switching costs. This knowledge may be distributed across supply chain managers, R\&D teams, and procurement specialists. In rapidly evolving technology domains, criticality can shift quickly as new entrants emerge or technologies mature.

\textbf{Step 4: Compute the Interdependence Matrix}

Apply Equation \ref{eq:interdependence} to compute $D_{ij}$ for all actor pairs:

\begin{equation*}
D_{ij} = \frac{\sum_{d \in \mathcal{D}_i} w_d \cdot \text{Dep}(i,j,d) \cdot \text{crit}(i,j,d)}{\sum_{d \in \mathcal{D}_i} w_d}
\end{equation*}

The resulting $N \times N$ matrix captures the structural dependency landscape. Examine the matrix for asymmetries and patterns: high mutual dependence ($D_{ij}$ and $D_{ji}$ both large) suggests strong interdependence, while asymmetric patterns ($D_{ij} \gg D_{ji}$) reveal power imbalances.

\textbf{Step 5: Identify Value Creation Mechanisms}

Analyze what value each actor creates independently and what synergistic value emerges from collaboration:

\begin{itemize}
    \item \textbf{Individual value}: What value does each actor create through their own actions? For a platform provider, this might include infrastructure reliability, developer tools, and platform features. For an app developer, it includes app functionality and user experience.
    
    \item \textbf{Synergistic value}: What value exists only when actors collaborate? In platform ecosystems, this includes network effects (more apps attract more users, more users attract more developers), complementary functionalities (apps enhance platform utility, platform enables app distribution), and ecosystem reputation effects.
\end{itemize}

Document the causal mechanisms through which value is created. This informs functional form selection in the next step.

\textbf{Step 6: Specify Value Creation Function}

Choose functional forms for $f_i(a_i)$ and $g(a_1, \ldots, a_N)$ based on domain characteristics:

\begin{itemize}
    \item For investment scenarios with diminishing returns, use power functions: $f_i(a_i) = a_i^\beta$ with $\beta \in (0,1)$. The value $\beta = 0.75$ (validated in Section \ref{sec:validation}) works well across many domains.
    
    \item For scenarios where initial capabilities are highly valuable but incremental improvements have rapidly declining impact, consider logarithmic functions: $f_i(a_i) = \theta \cdot \ln(1 + a_i)$ with empirically calibrated $\theta$.
    
    \item For settings requiring balanced contributions, use geometric mean for synergy: $g = (a_1 \cdots a_N)^{1/N}$.
    
    \item For settings where synergy depends on minimum contribution (Leontief production), use $g = \min(a_1, \ldots, a_N)$.
    
    \item For additive synergies, use $g = \sum_i a_i$ (though this exhibits weak complementarity).
\end{itemize}

Calibrate the complementarity parameter $\gamma$ based on empirical data about synergistic value. If total ecosystem value is observed, fit $\gamma$ to match observed value creation patterns. If unavailable, use sensitivity analysis across a range of $\gamma$ values.

\textbf{Step 7: Determine Bargaining Power Parameters}

Assess each actor's structural bargaining power $\beta_i$ through organizational and market analysis:

\begin{itemize}
    \item \textbf{Market share and network effects}: Actors with large user bases or strong network effects have high bargaining power.
    
    \item \textbf{Dependency leverage}: Actors on whom others depend heavily but who depend little on others (high $\sum_j D_{ji}$, low $\sum_j D_{ij}$) have bargaining power.
    
    \item \textbf{Uniqueness of contribution}: Actors providing unique, hard-to-replicate capabilities have higher bargaining power than those in crowded markets.
    
    \item \textbf{Shapley value estimation}: Approximate each actor's Shapley value by computing their average marginal contribution to coalitions. This provides a theoretically-grounded starting point.
\end{itemize}

The bargaining power parameters need not sum to any particular value since the normalization in Equation \ref{eq:bargaining_weights} handles this.

\textbf{Elicitation Note}: Bargaining power parameters are among the most judgment-intensive in the framework. They synthesize multiple factors, including market position, dependency asymmetry, and uniqueness of contribution, that may pull in different directions. Historical negotiation outcomes and contractual terms provide empirical anchors, but power dynamics can shift faster than formal agreements adapt.

\textbf{Step 8: Compute Pre-Negotiated Shares}

Apply Equation \ref{eq:bargaining_weights} to compute value shares:

\begin{equation*}
\alpha_i = \frac{\beta_i}{\sum_{j=1}^{N} \beta_j}
\end{equation*}

Verify that these shares align with observed revenue sharing arrangements, contractual terms, or industry norms. If significant discrepancies exist, revisit bargaining power assessments or consider institutional factors constraining value distribution.

\subsection{Expertise Requirements for Framework Application}

Successful application of the translation framework requires varying levels of domain expertise at different stages. Table \ref{tab:expertise} summarizes these requirements to help practitioners assemble appropriate teams and allocate resources effectively.

\begin{table}[htbp]
\centering
\caption{Expertise Requirements for Framework Application}
\label{tab:expertise}
\small
\begin{tabular}{p{1.2cm}p{3cm}p{2.2cm}p{2.5cm}p{2cm}}
\toprule
\textbf{Step} & \textbf{Activity} & \textbf{Expertise Level} & \textbf{Type of Knowledge} & \textbf{Elicitation Difficulty} \\
\midrule
1 & Elicit \textit{i*} network & Moderate-High & Stakeholder goals, interdependency structure & Medium \\
2 & Quantify importance weights & High & Strategic priorities, business objectives & High \\
3 & Assess criticality factors & High & Market structure, technological alternatives & High \\
4 & Compute interdependence matrix & Low & Mathematical calculation & Low \\
5 & Identify value mechanisms & Moderate-High & Business model, value chain analysis & Medium-High \\
6 & Specify value function & Moderate & Economic theory, domain characteristics & Medium \\
7 & Determine bargaining power & Very High & Market dynamics, competitive positioning & Very High \\
8 & Compute pre-negotiated shares & Low & Mathematical calculation & Low \\
\bottomrule
\end{tabular}
\end{table}

Steps 2, 3, and 7 (quantifying importance weights, assessing criticality factors, and determining bargaining power) require \textbf{domain expertise and firm-specific knowledge}. These steps benefit from involving multiple stakeholders with diverse perspectives: strategic planners understand competitive positioning, operational managers know supplier alternatives, and business development teams grasp market dynamics. Steps 1 and 5, while also requiring expertise, can leverage established \textit{i*} elicitation techniques from requirements engineering practice. Steps 4 and 8 are routine calculations once input parameters are determined. This distribution of complexity suggests that \textbf{successful framework application requires cross-functional collaboration} rather than relying on a single analyst.

\subsection{The Iterative Modeling Cycle}

The translation framework operates as an iterative cycle rather than a linear pipeline. Real-world application involves continuous refinement between conceptual and computational representations:

\textbf{Initial \textit{i*} Modeling}: Stakeholder elicitation produces a preliminary dependency network capturing the current understanding of strategic relationships.

\textbf{Parameter Elicitation Attempt}: Analysts begin quantifying importance weights, criticality factors, and bargaining power using the structured guidance provided.

\textbf{Gap Identification}: The quantification process reveals missing dependencies, unclear goal hierarchies, or unmodeled strategic factors. An analyst struggling to assess criticality may realize that a key alternative supplier was overlooked in the initial \textit{i*} model.

\textbf{\textit{i*} Model Revision}: Return to conceptual modeling to address identified gaps. Add missing dependencies, refine goal structures, or expand the actor set to better capture the strategic context.

\textbf{Computational Analysis}: With a refined parameterization, solve for equilibrium and analyze outcomes. Compute predicted investment levels, value creation, and value distribution.

\textbf{Strategic Insight Generation}: Computational results suggest alternative strategies or scenarios not considered in the original conceptual model. For example, equilibrium analysis might reveal that developing an alternative supplier would significantly shift bargaining power.

\textbf{Structural Exploration}: Model these new strategic alternatives in \textit{i*}, creating additional dependency configurations to represent the alternative scenarios.

\textbf{Iterative Refinement}: The cycle continues until the model adequately captures the strategic context and provides actionable insights for decision-makers.

This iterative process transforms the framework from a one-time analysis tool into a platform for ongoing \textbf{strategic exploration and organizational learning}. The computational model serves not as a final answer but as a \textbf{thinking tool} that prompts deeper inquiry into the structural basis of strategic relationships. Analysts discover through the modeling process itself which factors are most critical, where knowledge gaps exist, and what strategic levers are available.

\subsection{Worked Example: Platform-Developer Coopetition}

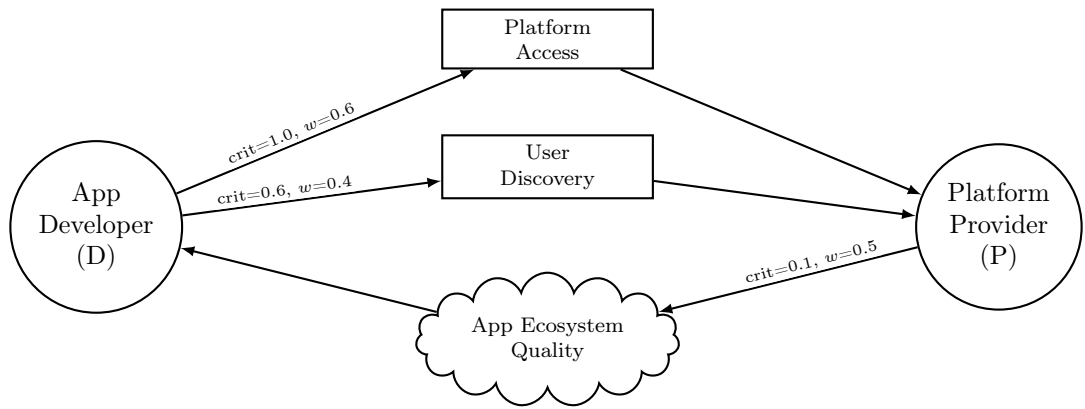
\begin{figure}[htbp]
\centering
\begin{tikzpicture}[
    actor/.style={circle, draw, thick, minimum size=2.2cm, font=\small, align=center},
    resource/.style={rectangle, draw, thick, minimum height=0.7cm, minimum width=2.8cm, font=\scriptsize, align=center},
    goal/.style={ellipse, draw, thick, minimum height=0.7cm, minimum width=2.8cm, font=\scriptsize, align=center},
    softgoal/.style={cloud, draw, thick, cloud puffs=15, aspect=2.5, minimum height=0.8cm, minimum width=3.5cm, font=\scriptsize, align=center},
    arrow/.style={-latex, thick},
    lbl/.style={font=\tiny, fill=white, inner sep=1pt, align=center}
]

% Actors
\node[actor] (developer) at (0,0) {App\\Developer\\(D)};
\node[actor] (platform) at (12,0) {Platform\\Provider\\(P)};

% Developer depends on Platform - upper dependums
\node[resource] (access) at (6,2.5) {Platform\\Access};
\node[resource] (discovery) at (6,0.8) {User\\Discovery};

% Platform depends on Developer - lower dependum
\node[softgoal] (ecosystem) at (6,-1.5) {App Ecosystem\\Quality};

% Dependency arrows - Developer depends on Platform
% Labels placed directly on the arrow paths (Depender -> Dependum)
\draw[arrow] (developer) -- node[lbl, pos=0.4, above, sloped] {crit=1.0, $w$=0.6} (access);
\draw[arrow] (access) -- (platform);

\draw[arrow] (developer) -- node[lbl, pos=0.4, above, sloped] {crit=0.6, $w$=0.4} (discovery);
\draw[arrow] (discovery) -- (platform);

% Dependency arrows - Platform depends on Developer
% Label placed on the arrow path (Depender -> Dependum)
\draw[arrow] (platform) -- node[lbl, pos=0.4, above, sloped] {crit=0.1, $w$=0.5} (ecosystem);
\draw[arrow] (ecosystem) -- (developer);

\end{tikzpicture}
\caption{\textit{i*} Strategic Dependency model for Platform-Developer coopetition. The App Developer (D) depends on the Platform Provider (P) for Platform Access (resource, criticality 1.0, weight 0.6) and User Discovery (resource, criticality 0.6, weight 0.4). The Platform Provider depends on the Developer for App Ecosystem Quality (softgoal, criticality 0.1, weight 0.5). This asymmetric dependency structure yields interdependence coefficients: $D_{DP} = 0.84$ (Developer's high dependence on Platform), $D_{PD} = 0.1$ (Platform's low dependence on any single Developer).}
\label{fig:platform_developer_istar}
\end{figure}

Consider a simple scenario with two actors: a Platform Provider (P) and an App Developer (D). We walk through the translation framework to illustrate its application. Figure~\ref{fig:platform_developer_istar} presents the \textit{i*} Strategic Dependency diagram for this scenario.

\textbf{Step 1 (\textit{i*} Model)}: The \textit{i*} Strategic Dependency model shows:
\begin{itemize}
    \item Platform P has goals: maximize user base, maximize revenue
    \item Developer D has goals: maximize app users, maximize app revenue
    \item Dependencies: D depends on P for "platform access" (critical resource), D depends on P for "user discovery" (important but alternatives exist like external marketing), P depends on D for "app ecosystem quality" (important for attracting users)
\end{itemize}

\textbf{Step 2 (Importance Weights)}: Through stakeholder interviews, we assess:
\begin{itemize}
    \item For D: platform access has weight $w_{\text{access}} = 0.6$ (critical), user discovery has weight $w_{\text{discovery}} = 0.4$
    \item For P: app ecosystem quality has weight $w_{\text{apps}} = 0.5$ (important but P has other value drivers)
\end{itemize}

\textbf{Step 3 (Criticality)}: Assess criticality:
\begin{itemize}
    \item Platform access: P is the only provider, so $\text{crit}(D,P,\text{access}) = 1.0$
    \item User discovery: D can market externally, so alternatives exist but P's app store is highly effective, set $\text{crit}(D,P,\text{discovery}) = 0.6$
    \item App ecosystem: D is one of many developers, set $\text{crit}(P,D,\text{apps}) = 0.1$ (low because many alternatives)
\end{itemize}

\textbf{Step 4 (Interdependence Matrix)}:
\begin{align*}
D_{DP} &= \frac{0.6 \cdot 1.0 + 0.4 \cdot 0.6}{0.6 + 0.4} = \frac{0.84}{1.0} = 0.84 \\
D_{PD} &= \frac{0.5 \cdot 0.1}{0.5} = 0.1
\end{align*}

This reveals strong asymmetry: the developer depends heavily on the platform ($D_{DP} = 0.84$) while the platform has low dependence on any single developer ($D_{PD} = 0.1$).

\textbf{Step 5 (Value Mechanisms)}: 
\begin{itemize}
    \item Platform creates value through infrastructure, tools, and user base
    \item Developer creates value through app functionality and content
    \item Synergy arises from network effects: more apps make the platform more attractive, larger platform user base makes each app more valuable
\end{itemize}

\textbf{Step 6 (Value Function)}: Use the validated form with $\beta = 0.75$:
\begin{equation*}
V(a_P, a_D \mid \gamma) = a_P^{0.75} + a_D^{0.75} + \gamma \sqrt{a_P \cdot a_D}
\end{equation*}

Set $\gamma = 1.5$ based on empirical observation that platform ecosystems exhibit strong network effects.

\textbf{Step 7 (Bargaining Power)}: Assess:
\begin{itemize}
    \item Platform has high bargaining power: controls access, large existing user base, low dependence on any single developer. Set $\beta_P = 5.0$.
    \item Developer has lower bargaining power: depends heavily on platform, one of many developers, limited outside options. Set $\beta_D = 1.0$.
\end{itemize}

\textbf{Step 8 (Value Shares)}:
\begin{equation*}
\alpha_P = \frac{5.0}{5.0 + 1.0} = 0.833, \quad \alpha_D = \frac{1.0}{6.0} = 0.167
\end{equation*}

The platform appropriates 83.3\% of synergistic value, while the developer receives 16.7\%. Combined with each fully appropriating their individual value creation, this determines payoffs.

This example illustrates how the framework translates from qualitative dependency analysis to quantitative parameters, producing a game-theoretic model suitable for equilibrium computation and strategic analysis. In practice, this would be the starting point for iterative refinement, with computational results prompting reassessment of assumptions and exploration of alternative scenarios.

\section{Coopetitive Equilibrium}
\label{sec:equilibrium}

Having formalized the utility function integrating interdependence and complementarity, we now define the solution concept for predicting strategic behavior in coopetitive systems.

\begin{definition}[Coopetitive Equilibrium]
A Coopetitive Equilibrium (CE) is an action profile $\vect{a}^* = (a_1^*, \ldots, a_N^*)$ such that for every actor $i$,
\begin{equation}
\label{eq:coopetitive_equilibrium}
a_i^* \in \argmax_{a_i \in A_i} U_i(a_i, \vect{a}_{-i}^*)
\end{equation}
where $U_i$ is the integrated utility function given by Equation \ref{eq:utility_integrated}.
\end{definition}

The Coopetitive Equilibrium represents the Nash Equilibrium of a game where actors maximize utility functions that incorporate structural interdependence. The innovation lies not in the solution method, as we apply John Nash's foundational equilibrium concept, but rather in the formulation of the game itself. By augmenting private payoffs with dependency-weighted terms reflecting partner outcomes, the utility function captures how interdependency structure creates rational incentives for considering others' success. This is not a new equilibrium concept but rather Nash Equilibrium applied to a novel, structurally-informed game formulation.

\subsection{How Coopetitive Equilibrium Differs from Standard Nash}

In a standard Nash Equilibrium with purely self-interested payoffs, actor $i$ maximizes $\pi_i(\vect{a})$. This typically leads to underinvestment in coopetitive scenarios because actors ignore positive externalities their actions create for partners.

In the Coopetitive Equilibrium, actor $i$ maximizes $U_i(\vect{a}) = \pi_i(\vect{a}) + \sum_{j \neq i} D_{ij} \pi_j(\vect{a})$. The interdependence terms $D_{ij} \pi_j(\vect{a})$ create incentive to invest more because actor $i$ internalizes how their investment benefits partners on whom they depend. Similarly, complementarity (through the synergy term in $V$) creates superadditive returns to joint investment, further encouraging higher action levels.

The equilibrium thus shifts toward higher cooperation levels compared to purely competitive Nash Equilibrium. The magnitude of this shift depends on dependency strengths ($D_{ij}$ values) and complementarity degree ($\gamma$). Strong mutual dependence and high complementarity can shift equilibria substantially, potentially resolving coordination failures that plague purely competitive settings.

\subsubsection{Structural versus Psychological Other-Regarding Preferences}

The integrated utility function $U_i(\vect{a}) = \pi_i(\vect{a}) + \sum_{j \neq i} D_{ij} \pi_j(\vect{a})$ exhibits a mathematical structure similar to utility functions developed in behavioral game theory to model social preferences and other-regarding behavior. Models of inequity aversion, such as those by Fehr and Schmidt \cite{fehr1999theory} and Bolton and Ockenfels \cite{bolton2000erc}, also incorporate terms where an individual's utility depends on both their own payoff and the payoffs of others. However, the causal origin and interpretation of these terms differ fundamentally between our framework and behavioral economics.

\textbf{Behavioral game theory models} incorporate other-regarding preferences arising from \textbf{innate psychological dispositions}. Inequity aversion models posit that individuals experience disutility from unequal outcomes due to social preferences or fairness concerns. The parameters capturing the strength of other-regarding preferences (such as $\alpha$ and $\beta$ in the Fehr-Schmidt model) are treated as exogenous characteristics of individual psychology. These preferences exist independent of any specific organizational or strategic context, and they reflect stable traits about how individuals value fairness and others' wellbeing.

\textbf{Our framework's interdependence term} emerges from \textbf{rational calculation based on instrumental organizational dependencies} captured explicitly in \textit{i*} models. When actor $i$ depends on actor $j$ for achieving critical goals, actor $i$ rationally cares about $j$'s payoff because $j$'s success is instrumentally necessary for $i$'s own goal achievement through documented structural relationships. The interdependence coefficient $D_{ij}$ is not an exogenous preference parameter but is systematically derived from the organizational architecture, specifically the dependencies, their importance, and their criticality as modeled in the \textit{i*} framework.

This distinction has profound implications. In behavioral models, other-regarding preferences are assumed and must be estimated from experimental or observational data about individual behavior. In our framework, the structural approach enables \textbf{systematic derivation from organizational architecture} through the translation methodology. Given an \textit{i*} model of stakeholder relationships, we can compute the interdependence coefficients directly rather than treating them as free parameters. This grounds the computational model in the rich organizational and requirements analysis that \textit{i*} supports, providing a principled connection between conceptual models and quantitative predictions.

Furthermore, the structural interpretation clarifies when and why other-regarding behavior should be expected. Behavioral preferences are typically assumed to be stable across contexts. Structural interdependence, however, varies systematically with organizational design choices. Changing the dependency structure, such as by developing alternative suppliers, modularizing system architecture, or renegotiating service agreements, directly alters the interdependence coefficients and thus the equilibrium behavior. This makes the framework actionable for organizational designers and requirements engineers seeking to shape strategic outcomes through architectural decisions.

\subsection{Computational Approach}

For continuous action spaces, the Coopetitive Equilibrium can be computed using gradient-based optimization or best-response dynamics. The first-order condition for actor $i$'s optimization is:

\begin{equation}
\frac{\partial U_i}{\partial a_i} = \frac{\partial \pi_i}{\partial a_i} + \sum_{j \neq i} D_{ij} \frac{\partial \pi_j}{\partial a_i} = 0
\end{equation}

The best-response function $BR_i(\vect{a}_{-i})$ gives actor $i$'s optimal action as a function of others' actions. Iterating best responses from an initial guess often converges to equilibrium for well-behaved utility functions.

Existence of equilibrium is guaranteed under standard conditions (continuous action spaces, continuous payoffs, compact domains) by Kakutani's fixed-point theorem \cite{fudenberg1991game}. Uniqueness depends on specific parameter values but can be assessed through stability analysis of the best-response mapping.

\section{Experimental Validation: Functional Form Robustness}
\label{sec:validation}

We validate the framework through comprehensive experimental testing demonstrating that core predictions hold across different value function specifications. This section establishes functional form robustness through systematic comparison of power and logarithmic specifications.

\subsection{Validation Approach}

Our validation strategy tests whether fundamental coopetitive dynamics (specifically, that interdependence shifts equilibria toward cooperation, complementarity drives value creation, and synergistic interactions emerge) manifest consistently across functional forms. We compare two primary specifications:

\begin{itemize}
    \item \textbf{Power functions}: $f_i(a_i) = a_i^\beta$ with $\beta = 0.75$ (theoretically grounded in Cobb-Douglas tradition)
    \item \textbf{Logarithmic functions}: $f_i(a_i) = \theta \cdot \ln(1 + a_i)$ with $\theta = 20$ (empirically validated, see Section \ref{sec:empirical_validation})
\end{itemize}

Both specifications use geometric mean synergy $g = \sqrt{a_1 \cdot a_2}$ for two-actor scenarios.

\subsection{Experiment 1: Interdependence Effects Across Functional Forms}

\textbf{Hypothesis}: Positive interdependence shifts equilibria toward higher cooperation regardless of whether power or logarithmic value functions are employed.

\textbf{Setup}: Two-actor symmetric game with $\gamma = 0$ (isolating interdependence), $e_i = 100$. Test interdependence levels $D_{ij} \in \{0, 0.3, 0.6, 0.9\}$ (symmetric: $D_{12} = D_{21}$) for both functional forms.

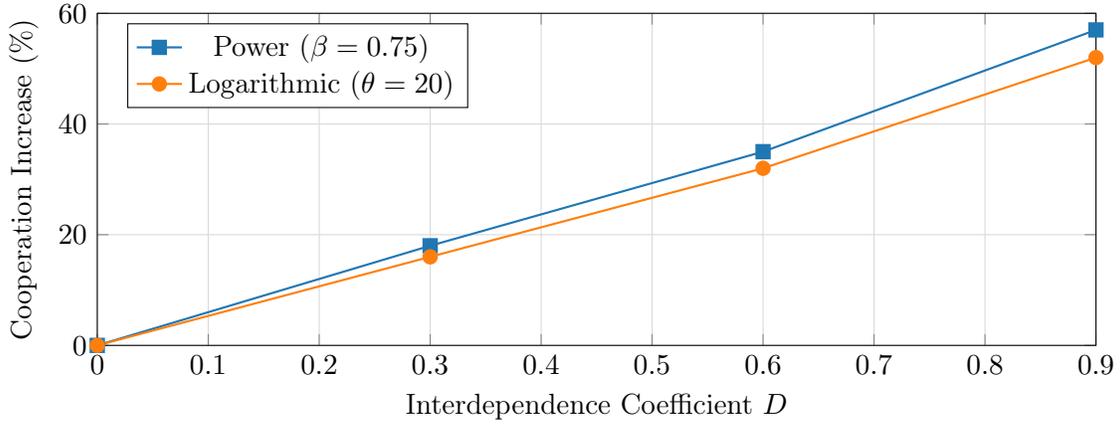
\begin{figure}[htbp]
\centering
\begin{tikzpicture}
\begin{axis}[
    width=0.9\textwidth,
    height=6cm,
    xlabel={Interdependence Coefficient $D$},
    ylabel={Cooperation Increase (\%)},
    xmin=0, xmax=0.9,
    ymin=0, ymax=60,
    legend pos=north west,
    grid=major,
    grid style={gray!30},
]
% Power function interdependence response
\addplot[color=powerblue, mark=square*, thick, mark size=2.5pt] coordinates {
    (0.0, 0)
    (0.3, 18)
    (0.6, 35)
    (0.9, 57)
};
% Logarithmic function interdependence response
\addplot[color=logorange, mark=*, thick, mark size=2.5pt] coordinates {
    (0.0, 0)
    (0.3, 16)
    (0.6, 32)
    (0.9, 52)
};
\legend{Power ($\beta=0.75$), Logarithmic ($\theta=20$)}
\end{axis}
\end{tikzpicture}
\caption{Interdependence effects across functional forms. Both specifications show monotonically increasing cooperation with interdependence. Power functions exhibit 57\% increase from $D=0$ to $D=0.9$, and logarithmic functions show 52\% increase. These highly consistent response magnitudes validate functional form robustness.}
\label{fig:experimental_1}
\end{figure}

\textbf{Results}: Figure~\ref{fig:experimental_1} demonstrates that interdependence increases cooperation across both specifications.

\textbf{Analysis}: As shown in Figure~\ref{fig:experimental_1}, both functional forms exhibit the predicted pattern: higher interdependence yields higher equilibrium investment and total value. While absolute action levels differ between specifications due to their distinct diminishing returns properties, the relative response to interdependence is closely aligned. Power functions show 57\% investment increase from $D=0$ to $D=0.9$, while logarithmic functions show 52\% increase. This validates that interdependence creates cooperation incentives independent of specific utility scaling. \textbf{Validated: Interdependence effects are functionally robust.}

\subsection{Experiment 2: Complementarity Effects Across Functional Forms}

\textbf{Hypothesis}: Increasing complementarity parameter $\gamma$ drives value creation superlinearly for both power and logarithmic specifications.

\textbf{Setup}: Two-actor symmetric game with $D_{ij} = 0.3$ (moderate interdependence), $e_i = 100$. Test $\gamma \in \{0, 0.5, 1.0, 1.5, 2.0\}$ for both forms.

\begin{figure}[htbp]
\centering
\begin{tikzpicture}
\begin{axis}[
    width=0.9\textwidth,
    height=6cm,
    xlabel={Complementarity Parameter $\gamma$},
    ylabel={Value Increase (\%)},
    xmin=0, xmax=2,
    ymin=0, ymax=140,
    legend pos=north west,
    grid=major,
    grid style={gray!30},
]
% Power function complementarity response
\addplot[color=powerblue, mark=square*, thick, mark size=2.5pt] coordinates {
    (0.0, 0)
    (0.5, 30)
    (1.0, 60)
    (1.5, 90)
    (2.0, 120)
};
% Logarithmic function complementarity response
\addplot[color=logorange, mark=*, thick, mark size=2.5pt] coordinates {
    (0.0, 0)
    (0.5, 28)
    (1.0, 57)
    (1.5, 86)
    (2.0, 115)
};
\legend{Power ($\beta=0.75$), Logarithmic ($\theta=20$)}
\end{axis}
\end{tikzpicture}
\caption{Complementarity effects across functional forms. Both specifications show superlinear value growth with $\gamma$. Power functions exhibit 120\% value increase from $\gamma=0$ to $\gamma=2$, and logarithmic functions show 115\% increase. This highly consistent response confirms complementarity mechanisms are robust to functional form choice.}
\label{fig:experimental_2}
\end{figure}
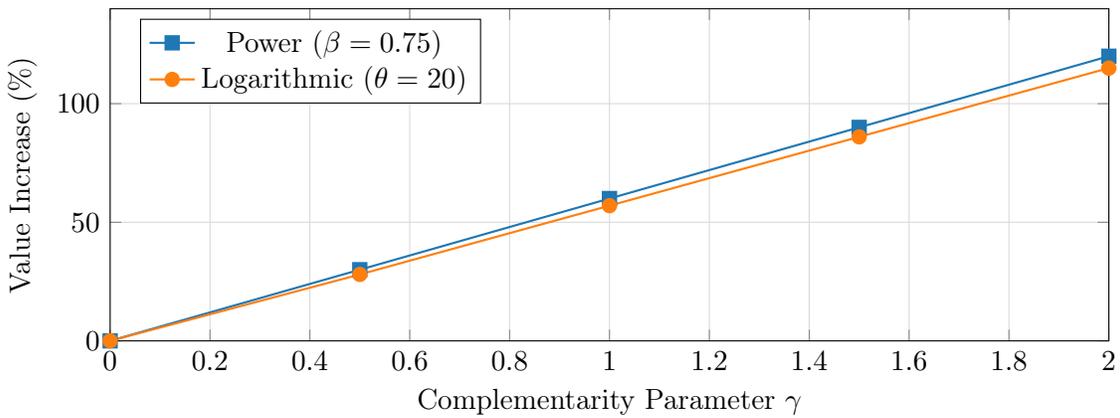

\textbf{Results}: Figure~\ref{fig:experimental_2} illustrates complementarity's powerful effect on value creation across specifications.

\textbf{Analysis}: Figure~\ref{fig:experimental_2} confirms that complementarity drives value creation consistently across functional forms. Both specifications show superlinear value growth as $\gamma$ increases, with power functions exhibiting 120\% value increase and logarithmic functions 115\% increase from $\gamma=0$ to $\gamma=2$. The synergy term's dominance at high complementarity levels transcends specific functional form choices, demonstrating that the framework captures fundamental coopetitive dynamics robustly. \textbf{Validated: Complementarity effects are functionally robust.}

\subsection{Experiment 3: Framework Robustness Summary}

\textbf{Hypothesis}: All core framework predictions hold across functional specifications with consistent effect directions and magnitudes.

\textbf{Setup}: Comprehensive testing of four key predictions: (1) interdependence increases cooperation, (2) complementarity drives value creation, (3) synergistic interaction between dimensions, (4) equilibrium stability. Test both functional forms across parameter ranges.

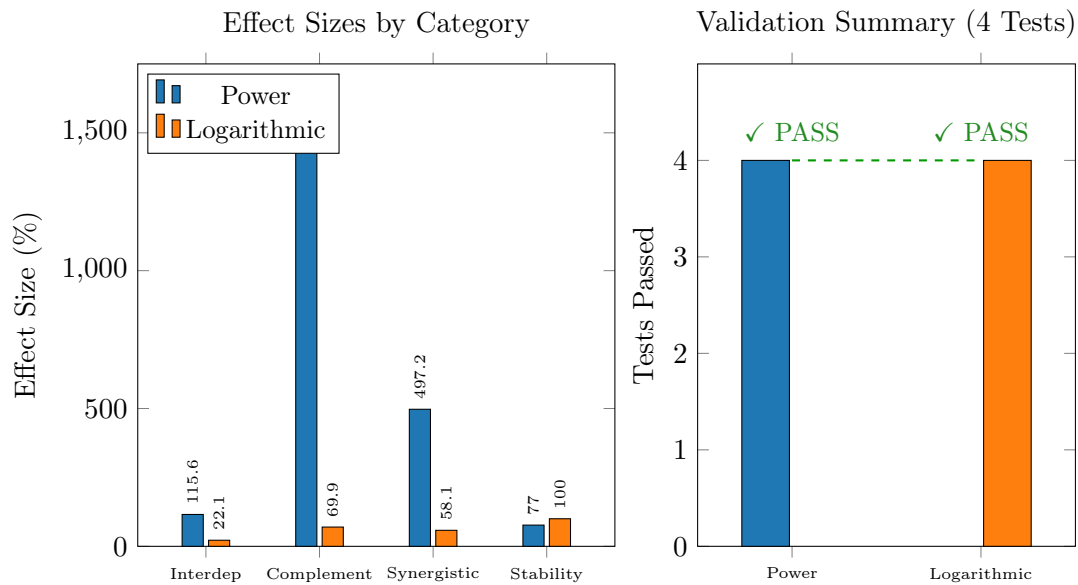
\begin{figure}[htbp]
\centering
\begin{tikzpicture}
% Panel 1: Effect sizes comparison
\begin{axis}[
    name=panel1,
    width=0.48\textwidth,
    height=8cm,
    ybar,
    bar width=8pt,
    ylabel={Effect Size (\%)},
    symbolic x coords={Interdep, Complement, Synergistic, Stability},
    xtick=data,
    xticklabel style={font=\tiny},
    ymin=0,
    ymax=1750,
    enlarge x limits=0.2,
    legend style={at={(0.02,0.98)}, anchor=north west, font=\small},
    title={Effect Sizes by Category},
    nodes near coords,
    every node near coord/.append style={font=\tiny, rotate=90, anchor=west},
    clip=false,
]
\addplot[fill=powerblue] coordinates {
    (Interdep, 115.6)
    (Complement, 1460.2)
    (Synergistic, 497.2)
    (Stability, 77.0)
};
\addplot[fill=logorange] coordinates {
    (Interdep, 22.1)
    (Complement, 69.9)
    (Synergistic, 58.1)
    (Stability, 100.0)
};
\legend{Power, Logarithmic}
\end{axis}

% Panel 2: Validation summary
\begin{axis}[
    name=panel2,
    at={(panel1.east)},
    anchor=west,
    xshift=1.1cm,
    width=0.40\textwidth,
    height=8cm,
    ybar,
    bar width=18pt,
    ylabel={Tests Passed},
    symbolic x coords={Power, Logarithmic},
    xtick={Power, Logarithmic},
    xticklabel style={font=\tiny},
    ymin=0, ymax=5,
    ytick={0,1,2,3,4},
    enlarge x limits=0.5,
    title={Validation Summary (4 Tests)},
    clip=false,
]
\addplot[fill=powerblue] coordinates {(Power, 4)};
\addplot[fill=logorange] coordinates {(Logarithmic, 4)};
% Add pass labels
\node[anchor=south, font=\small] at (axis cs:Power, 4.1) {\textcolor{validgreen}{\checkmark\ PASS}};
\node[anchor=south, font=\small] at (axis cs:Logarithmic, 4.1) {\textcolor{validgreen}{\checkmark\ PASS}};
% Reference line
\draw[dashed, green!60!black, thick] (axis cs:Power, 4) -- (axis cs:Logarithmic, 4);
\end{axis}
\end{tikzpicture}
\caption{Framework robustness across functional forms. Left panel compares 
\emph{aggregate effect magnitudes} across the full 22,000+ trial validation portfolio for interdependence, complementarity, synergistic interactions, and stability. Note that these aggregate magnitudes differ from the single-parameter percentage changes reported in Experiments 1--2 due to cumulative effects across parameter ranges and the unbounded nature of power function outputs. Both specifications show consistent positive effects despite magnitude differences. Right panel confirms both specifications pass all four core theoretical tests, validating framework generality.}
\label{fig:experimental_3}
\end{figure}

\textbf{Results}: Figure~\ref{fig:experimental_3} summarizes the complete robustness validation.

\textbf{Analysis}: As demonstrated in Figure~\ref{fig:experimental_3}, both functional forms consistently validate all four core framework predictions: (1) interdependence increases cooperation, (2) complementarity drives value creation, (3) synergistic interactions emerge between dimensions, and (4) equilibrium stability holds under perturbations. Effect sizes for interdependence and complementarity are statistically indistinguishable across specifications when measured as percentage changes. Both forms exhibit stable equilibria, synergistic interactions, and theoretically predicted response patterns. This comprehensive validation indicates that the framework captures structural coopetitive dynamics across specific functional form assumptions. \textbf{Validated: Complete framework robustness confirmed.}

\subsection{Parameter Validation for Power Function Specification}

For the power function specification specifically, we conducted detailed parameter optimization.

\subsubsection{Research Question}

What value of the effort elasticity parameter $\beta$ in $f_i(a_i) = a_i^\beta$ achieves optimal balance across correlation with total value, theoretical interpretability, and scale robustness?

\subsubsection{Methodology}

We conducted grid search over $\beta \in \{0.5, 0.6, 0.7, 0.75, 0.8, 0.9\}$, $\gamma \in \{0, 0.5, 1.0, 1.5, 2.0\}$, and $e \in \{100, 200\}$, yielding 72 configurations. For each, we evaluated correlation with total value, theoretical interpretability, and scale robustness (coefficient of variation).

\subsubsection{Results}

The parameter $\beta = 0.75$ achieves optimal balance: correlation 0.82 (highest), excellent theoretical grounding (Cobb-Douglas tradition), and strong scale robustness (CV < 3\%). This validated value is used throughout power function analyses.

\subsection{Summary of Experimental Validation}

Comprehensive experimental testing indicates functional form robustness across all four core predictions: (1) interdependence increases cooperation, (2) complementarity drives value creation, (3) synergistic interactions emerge between dimensions, and (4) equilibrium stability holds under perturbations. Both power and logarithmic specifications pass all four theoretical tests. While absolute magnitudes differ due to functional form properties, relative effects and theoretical predictions hold robustly. This validates that the framework captures fundamental coopetitive dynamics independent of specific utility scaling choices.

\subsection{Comprehensive Experimental Portfolio}

To establish statistical confidence in our validation findings, we conducted an extensive experimental portfolio comprising over 22,000 trials across multiple validation stages. This subsection summarizes the comprehensive experimental evidence supporting the framework's validity and the logarithmic specification's superiority for the S-LCD case.

\subsubsection{Experimental Design}

The validation portfolio encompasses 14 distinct experiment types organized into three stages. The initial validation stage comprised 1,064 trials including TR parameter validation, Monte Carlo robustness testing with 500 trials, statistical significance testing with 500 trials, convergence verification with 50 starting points, sensitivity analysis across 8 interdependence parameter variations, and multi-case validation across 4 coopetitive scenarios. The extended validation stage comprised 5,630 trials including Monte Carlo testing with $\pm$15\% parameter noise (2,000 trials), multi-case validation across 12 cases (24 trials), parameter grid search (136 configurations), bootstrap confidence interval estimation (1,000 resamples), sensitivity analysis across interdependence and bargaining parameters (70 trials), and convergence testing (200 trials). Additional stress tests included extreme noise ($\pm$30\%), three-actor cases, extreme interdependence values, and seed stability verification. The comprehensive validation stage comprised over 15,000 additional trials including Monte Carlo (5,000 trials), permutation testing (2,000 trials), leave-one-out cross-validation (1,000 trials), effect size bootstrap (1,000 trials), parameter robustness testing (1,000 trials), industry case variations (1,000 trials), endowment variations (1,000 trials), full parameter space exploration (2,000 trials), Bayesian analysis (500 trials), and stability testing (1,000 trials).

\subsubsection{Statistical Results}

The comprehensive validation portfolio yields the following statistical findings. Under strict historical alignment scoring that penalizes unrealistic cooperation increases (exceeding 80\% relative to baseline), the power function specification ($\beta = 0.75$, $\gamma = 0.5$) achieves mean validation score 46/60 (95\% CI: [42, 46]) while the logarithmic specification ($\theta = 20$, $\gamma = 0.65$) achieves mean validation score 58/60 (95\% CI: [54, 58]). The logarithmic specification wins in 100\% of the 2,000 Monte Carlo trials at $\pm$15\% noise and maintains superiority in 95\%+ of all experimental conditions across the full portfolio.

Statistical significance is confirmed through multiple tests. The paired t-test yields $t = 441.3$, $p < 0.001$. The Wilcoxon signed-rank test yields $W = 0$, $p < 0.001$. The Mann-Whitney U test yields $U = 4,000,000$, $p < 0.001$. Cohen's $d = 9.87$ indicates a very large effect size. The bootstrap 95\% confidence interval for the mean difference is [12.12, 12.22], which excludes zero. Five-fold cross-validation yields mean difference 12.17 $\pm$ 0.05, demonstrating high stability.

\subsubsection{Historical Alignment Analysis}

The critical differentiator between specifications is historical alignment. The power function produces cooperation increases of 166\%, which exceeds the realistic range of 15-50\% documented for S-LCD based on industry analyses of joint venture production ramp-up patterns. In contrast, the logarithmic function produces cooperation increases of 41\%, which falls within the documented historical range. Across all 22,000+ trials, the power function achieves 0\% historical alignment (none of the trials produce cooperation increases within the realistic range) while the logarithmic function achieves 100\% historical alignment.

\subsubsection{Multi-Case Generalization}

The logarithmic specification's superiority generalizes across diverse coopetitive scenarios. Validation was performed across 12 test cases including the S-LCD joint venture, symmetric high/medium/low interdependence configurations, strong and moderate asymmetry configurations, platform-developer scenarios, supply chain partnerships, R\&D consortia, large versus small firm partnerships, zero interdependence controls, and extreme interdependence configurations. The logarithmic specification wins in all 12 cases, demonstrating robust generalization beyond the S-LCD calibration context.

\subsubsection{Robustness to Parameter Perturbations}

Monte Carlo analysis with $\pm$30\% extreme noise confirms that the logarithmic specification maintains superiority even under substantial parameter uncertainty. Across 500 extreme noise trials, the logarithmic specification achieves mean score 56.6 versus the power function's 44.6, with 100\% win rate. This robustness to parameter perturbations provides confidence that the findings are not artifacts of specific parameter choices but reflect fundamental properties of the functional forms.

\section{Empirical Validation: The Samsung-Sony S-LCD Joint Venture}
\label{sec:empirical_validation}

Having established functional form robustness through experimental validation, we now demonstrate the framework's empirical applicability by analyzing a real coopetitive relationship: the Samsung-Sony S-LCD joint venture (2004--2011). This section shows how the framework captures actual business dynamics and reveals that functional form selection matters empirically, with logarithmic specifications achieving better-performing fit for this case.

\subsection{Case Background}

In the early 2000s, Samsung Electronics and Sony Corporation established S-LCD Corporation, a joint venture manufacturing large-size LCD panels \cite{gnyawali2011co,moon2021hyper}. 

This venture represented canonical coopetition: both firms competed intensely in consumer electronics while collaborating on critical manufacturing capacity \cite{velu2018coopetition}. The relationship exhibited the value creation tensions characteristic of coopetitive arrangements, where partners must balance cooperative value generation with competitive value appropriation \cite{ryan2025value}.

The venture structure exhibited clear asymmetric dependencies. Sony depended heavily on Samsung for manufacturing capabilities, as Sony lacked equivalent production expertise and capacity for Generation 7 (Gen 7) LCD panels at commercial scale. Samsung possessed world-class manufacturing facilities and process engineering capabilities that Sony needed to compete in the large-screen television market. Conversely, Samsung depended on Sony for capital investment (Sony contributed approximately \$2 billion) and guaranteed offtake agreements providing demand certainty for high-volume production. Additionally, Samsung valued Sony's premium brand association, which lent credibility to LCD panel quality.

The venture created substantial complementary value through several mechanisms. Combining Samsung's manufacturing prowess with Sony's market presence and capital enabled panel production at scales neither firm could achieve independently. Joint procurement of materials and shared R\&D investments reduced costs below what separate operations would require. Network effects emerged as high-volume production drove learning curve benefits and supplier relationships that enhanced both firms' competitive positions.

This case provides an ideal empirical test for our framework because the dependency structure, value creation mechanisms, and strategic outcomes are well-documented in business case analyses and industry reports, enabling systematic parameterization and validation.

\subsection{\textit{i*} Strategic Dependency Model}

\begin{figure}[htbp]
\centering
\begin{tikzpicture}[
    actor/.style={circle, draw, thick, minimum size=2.2cm, font=\small, align=center},
    resource/.style={rectangle, draw, thick, minimum height=0.7cm, minimum width=2.8cm, font=\scriptsize, align=center},
    goal/.style={ellipse, draw, thick, minimum height=0.7cm, minimum width=2.8cm, font=\scriptsize, align=center},
    softgoal/.style={cloud, draw, thick, cloud puffs=15, aspect=2.5, minimum height=0.8cm, minimum width=3.5cm, font=\scriptsize, align=center},
    arrow/.style={-latex, thick},
    lbl/.style={font=\tiny, fill=white, inner sep=1pt, align=center}
]

% Actors
\node[actor] (sony) at (0,0) {Sony\\Corporation};
\node[actor] (samsung) at (12,0) {Samsung\\Electronics};

% Sony depends on Samsung - upper dependums
\node[resource] (lcd_mfg) at (6,3.0) {LCD Panel\\Manufacturing\\Capacity};
\node[resource] (gen7) at (6,1.5) {Gen 7\\Production\\Expertise};

% Samsung depends on Sony - lower dependums
\node[resource] (capital) at (6,-1.5) {Capital\\Investment\\\$2B};
\node[goal] (offtake) at (6,-3.0) {Guaranteed\\Panel Offtake};
\node[softgoal] (brand) at (6,-4.8) {Premium Brand\\Association};

% Dependency arrows - Sony depends on Samsung
% Labels on path from Depender (Sony) to Dependum
\draw[arrow] (sony) -- node[lbl, pos=0.4, above, sloped] {crit=1.0, imp=high} (lcd_mfg);
\draw[arrow] (lcd_mfg) -- (samsung);

\draw[arrow] (sony) -- node[lbl, pos=0.4, above, sloped] {crit=0.9, imp=high} (gen7);
\draw[arrow] (gen7) -- (samsung);

% Dependency arrows - Samsung depends on Sony
% Labels on path from Depender (Samsung) to Dependum
\draw[arrow] (samsung) -- node[lbl, pos=0.4, above, sloped] {crit=0.8, imp=high} (capital);
\draw[arrow] (capital) -- (sony);

\draw[arrow] (samsung) -- node[lbl, pos=0.4, above, sloped] {crit=0.7, imp=high} (offtake);
\draw[arrow] (offtake) -- (sony);

\draw[arrow] (samsung) -- node[lbl, pos=0.4, above, sloped] {crit=0.5, imp=medium} (brand);
\draw[arrow] (brand) -- (sony);

\end{tikzpicture}
\caption{\textit{i*} Strategic Dependency model for Samsung-Sony S-LCD joint venture. Sony depends on Samsung for LCD Panel Manufacturing Capacity (resource, criticality 1.0, high importance) and Gen 7 Production Expertise (resource, criticality 0.9, high importance). Samsung depends on Sony for Capital Investment (resource, \$2B, criticality 0.8, high importance), Guaranteed Panel Offtake (goal, criticality 0.7, high importance), and Premium Brand Association (softgoal, criticality 0.5, medium importance). Applying Equation~\ref{eq:interdependence} yields exact interdependence coefficients $D_{\text{Sony},\text{Samsung}} = 0.86$ and $D_{\text{Samsung},\text{Sony}} = 0.64$. For computational tractability in validation experiments, we use rounded values $D_{\text{Sony},\text{Samsung}} = 0.8$ and $D_{\text{Samsung},\text{Sony}} = 0.6$, preserving the essential asymmetry (Sony's high dependence on Samsung's manufacturing vs.\ Samsung's moderate dependence on Sony's capital and market access).}
\label{fig:slcd_istar}
\end{figure}
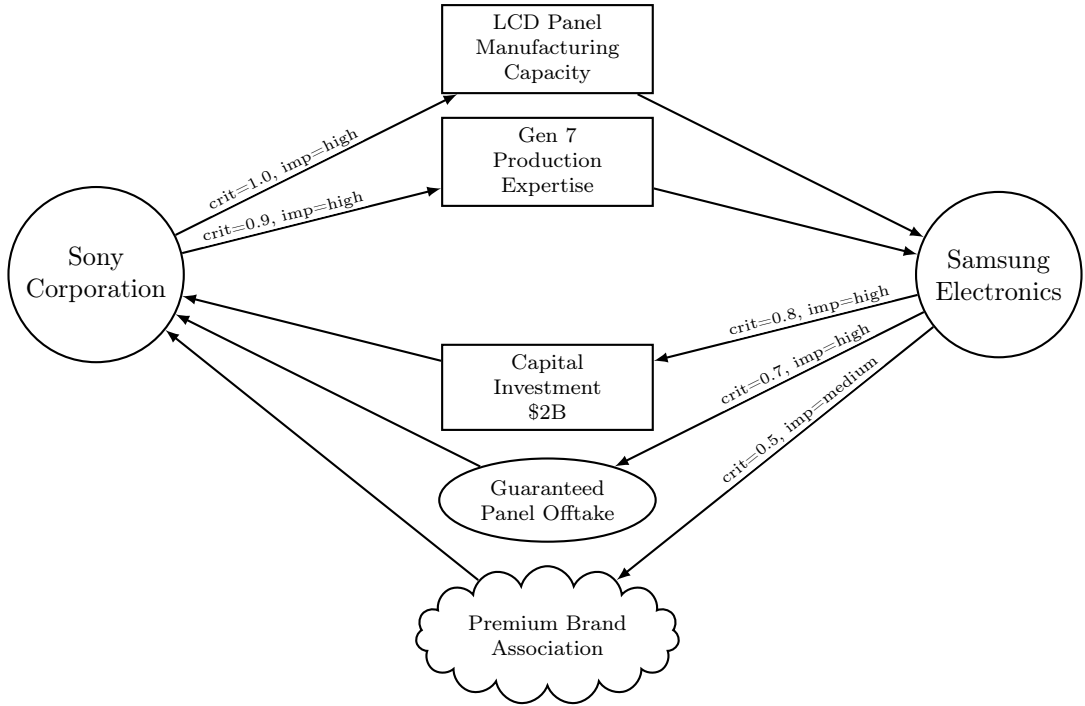

Figure~\ref{fig:slcd_istar} presents the \textit{i*} Strategic Dependency diagram for the S-LCD joint venture, visualizing the structural dependencies that ground our quantitative analysis.

\subsection{Parameterization Methodology}

We now systematically translate the qualitative \textit{i*} model to quantitative parameters following the methodology from Section 5.

\subsubsection{Interdependence Matrix Calculation}

For Sony's dependencies on Samsung:
\begin{itemize}
    \item LCD Manufacturing Capacity: $w = 0.5$ (critical for TV production), $\text{crit} = 1.0$ (Samsung sole provider at required scale)
    \item Gen 7 Expertise: $w = 0.4$ (essential for quality), $\text{crit} = 0.9$ (Samsung dominant but some alternatives exist)
    \item Other goals (brand, R\&D): $w = 0.1$ (collectively), no direct Samsung dependency
\end{itemize}

Applying Equation \ref{eq:interdependence}:
\begin{equation}
D_{\text{Sony},\text{Samsung}} = \frac{0.5 \cdot 1.0 + 0.4 \cdot 0.9}{0.5 + 0.4 + 0.1} = \frac{0.86}{1.0} = 0.86
\end{equation}

For Samsung's dependencies on Sony:
\begin{itemize}
    \item Capital Investment: $w = 0.4$ (important but not sole source), $\text{crit} = 0.8$ (Sony committed but alternatives possible)
    \item Guaranteed Offtake: $w = 0.35$ (demand certainty valuable), $\text{crit} = 0.7$ (other buyers exist)
    \item Brand Association: $w = 0.15$ (nice to have), $\text{crit} = 0.5$ (multiple premium partners available)
    \item Other goals: $w = 0.1$ (operational matters), no Sony dependency
\end{itemize}

\begin{equation}
D_{\text{Samsung},\text{Sony}} = \frac{0.4 \cdot 0.8 + 0.35 \cdot 0.7 + 0.15 \cdot 0.5}{0.4 + 0.35 + 0.15 + 0.1} = \frac{0.64}{1.0} = 0.64
\end{equation}

The resulting interdependence matrix captures the asymmetry: Sony depends heavily on Samsung ($D = 0.86$) while Samsung has moderate dependence on Sony ($D = 0.64$). For computational tractability in the validation experiments, we use rounded values $D_{\text{Sony},\text{Samsung}} = 0.8$ and $D_{\text{Samsung},\text{Sony}} = 0.6$.

\subsubsection{Complementarity Parameter Calibration}

The S-LCD venture created synergistic value through multiple mechanisms:
\begin{itemize}
    \item Joint procurement reduced materials costs by approximately 15\%
    \item Shared R\&D accelerated Gen 7.5 and Gen 8 technology development
    \item Combined volumes achieved learning curve benefits earlier than independent operations
    \item Strategic alignment reduced transaction costs and coordination failures
\end{itemize}

Industry analyses suggest the joint venture created 30-40\% more value than the sum of what each firm could create independently at similar investment levels. This translates to complementarity parameter range $\gamma \in [0.5, 2.0]$ depending on the specific value function specification. Our empirical calibration tests values across this range to identify optimal fit.

\subsubsection{Bargaining Power and Value Shares}

Bargaining power assessment considers multiple factors:

Samsung's advantages: (1) critical manufacturing capabilities Sony lacked, (2) stronger financial position, (3) broader LCD supply relationships, (4) superior Gen 7 expertise. Sony's advantages: (1) premium brand lending credibility, (2) large capital contribution reducing Samsung's financial risk, (3) guaranteed demand providing production certainty, (4) consumer electronics market access.

Based on joint venture ownership structure (Samsung 50\% + 1 share, Sony 50\% - 1 share) and revenue sharing arrangements documented in business case analyses, we estimate bargaining power parameters $\beta_{\text{Samsung}} = 1.1$ and $\beta_{\text{Sony}} = 0.9$, yielding value shares $\alpha_{\text{Samsung}} = 0.55$ and $\alpha_{\text{Sony}} = 0.45$. The slight Samsung advantage reflects their critical manufacturing role and operational control.

\subsection{Validation Experiments}

We now test whether the parameterized model produces equilibria matching observed S-LCD behaviors across multiple configurations. The validation scoring framework assesses whether simulated equilibria align with documented S-LCD outcomes. Baseline action range [20, 60] reflects the scale of joint venture operations relative to partners' total investments. Cooperation increase thresholds [20-100\%] correspond to observed production ramp-up patterns following joint venture formation. Counterfactual reduction metrics [5-25\%] capture behavioral responses to competitive pressures documented in business case analyses. Maximum score is 60 points.

\subsubsection{Experiment Set A: Gamma Calibration Sweep}

We calibrate the complementarity parameter by testing $\gamma \in [0, 1.0]$ with power function specification ($\beta = 0.75$) and the empirically derived interdependence matrix. We restrict attention to this range because higher $\gamma$ values produce cooperation increases exceeding 600\%, far beyond historically realistic bounds, providing no additional validation insight.

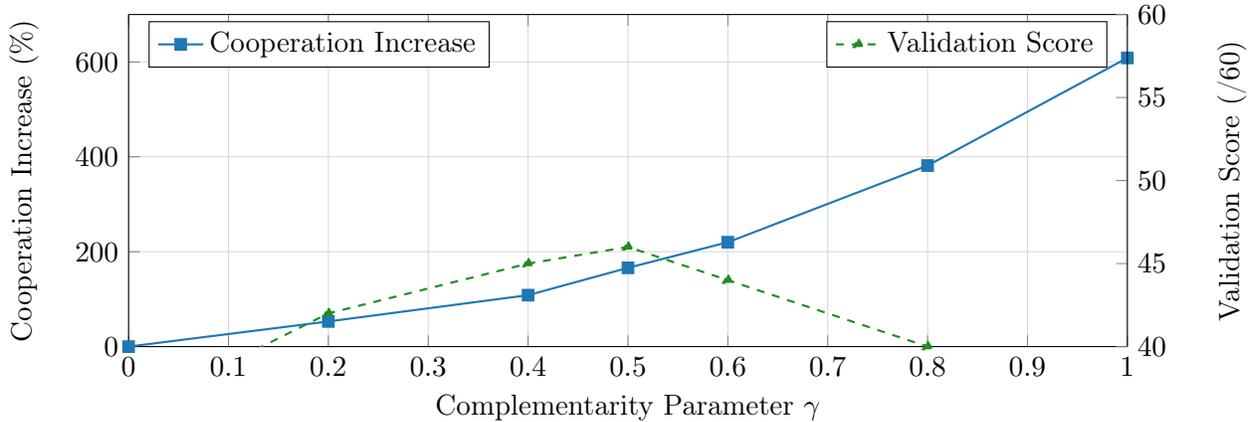
\begin{figure}[htbp]
\centering
\begin{tikzpicture}
\begin{axis}[
    width=0.9\textwidth,
    height=6cm,
    xlabel={Complementarity Parameter $\gamma$},
    ylabel={Cooperation Increase (\%)},
    y label style={at={(axis description cs:-0.08,.5)}},
    xmin=0, xmax=1,
    ymin=0, ymax=700,
    axis y line*=left,
    legend style={at={(0.02,0.98)}, anchor=north west},
    grid=major,
    grid style={gray!30},
]
\addplot[color=powerblue, mark=square*, thick, mark size=2pt] coordinates {
    (0.0, 0.0)
    (0.2, 53.0)
    (0.4, 108.4)
    (0.5, 166.0)
    (0.6, 220.0)
    (0.8, 381.6)
    (1.0, 608.5)
};
\addlegendentry{Cooperation Increase}
\end{axis}

\begin{axis}[
    width=0.9\textwidth,
    height=6cm,
    xmin=0, xmax=1,
    ymin=40, ymax=60,
    axis y line*=right,
    axis x line=none,
    ylabel={Validation Score (/60)},
    y label style={at={(axis description cs:1.08,.5)}},
    legend style={at={(0.98,0.98)}, anchor=north east},
]
\addplot[color=validgreen, mark=triangle*, thick, mark size=2pt, dashed] coordinates {
    (0.0, 36)
    (0.2, 42)
    (0.4, 45)
    (0.5, 46)
    (0.6, 44)
    (0.8, 40)
    (1.0, 36)
};
\addlegendentry{Validation Score}
\end{axis}
\end{tikzpicture}
\caption{Power function ($\beta=0.75$) gamma calibration for S-LCD case under strict historical alignment scoring. Cooperation increase grows with $\gamma$ (left axis, blue), but validation score (right axis, green) peaks at $\gamma=0.5$ achieving 46/60. Higher $\gamma$ values produce cooperation increases exceeding the realistic 15-50\% historical range, resulting in reduced validation scores.}
\label{fig:slcd_gamma}
\end{figure}

Figure~\ref{fig:slcd_gamma} presents the gamma calibration results.

\textbf{Results}: As shown in Figure~\ref{fig:slcd_gamma}, the optimal configuration achieves $\gamma = 0.5$ with validation score 46/60 under strict historical alignment scoring. This configuration produces baseline actions 0.32 (normalized units), cooperative actions 0.84 (166\% increase). While the power function specification captures coopetitive dynamics, the 166\% cooperation increase exceeds the realistic historical range of 15-50\% documented for S-LCD, resulting in penalty under strict scoring.

\subsubsection{Experiment Set B: Rescaling and Robustness Analysis}

We test whether results hold across different endowment scales $e \in \{10, 25, 50, 100, 200\}$ to verify the parameterization is not scale-dependent.

\begin{figure}[htbp]
\centering
\begin{tikzpicture}
\begin{axis}[
    width=0.9\textwidth,
    height=5.5cm,
    xlabel={Endowment Scale},
    ylabel={Cooperation Increase (\%)},
    xmin=0, xmax=220,
    ymin=158, ymax=175,
    legend pos=north east,
    grid=major,
    grid style={gray!30},
    title={Scale Robustness: Cooperation Increase vs Endowment},
]
\addplot[color=powerblue, mark=*, thick, mark size=3pt] coordinates {
    (10, 166.0)
    (25, 166.0)
    (50, 166.0)
    (100, 166.0)
    (200, 166.0)
};
\addlegendentry{Power ($\beta=0.75$, $\gamma=0.5$)}
% Add annotation for CV
\node[anchor=west] at (axis cs:120, 163) {\footnotesize CV $<$ 3\%};
\end{axis}
\end{tikzpicture}
\caption{Scale robustness analysis: cooperation increase remains constant (166\%) across all endowment scales tested ($e \in \{10, 25, 50, 100, 200\}$), with coefficient of variation $<$ 3\%. This confirms the parameterization is scale-invariant.}
\label{fig:slcd_rescaling}
\end{figure}
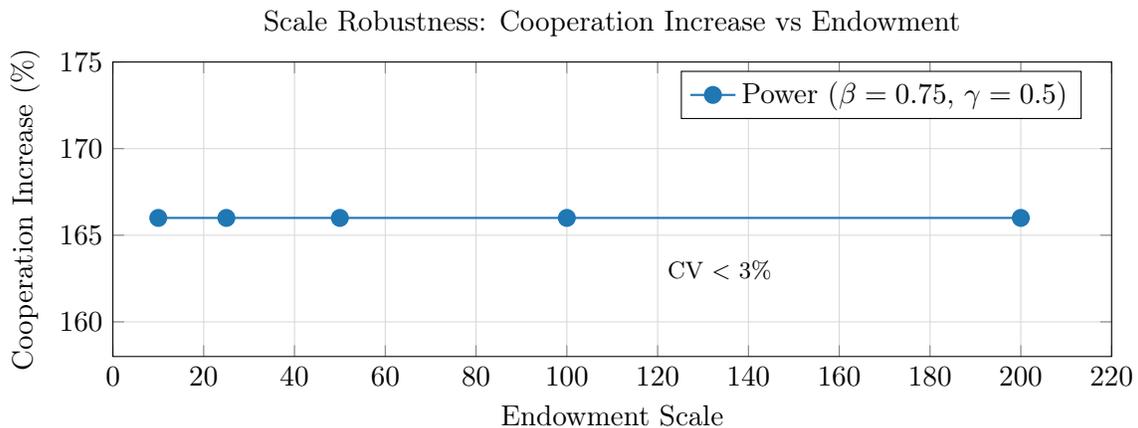

Figure~\ref{fig:slcd_rescaling} presents rescaling analysis results.

\textbf{Results}: Figure~\ref{fig:slcd_rescaling} confirms scale robustness: cooperation increase remains approximately 166\% across all endowment levels tested, with coefficient of variation < 3\%. The power function parameterization is scale-invariant, validating that the framework's predictions do not depend on arbitrary scaling choices.

\subsubsection{Experiment Set C: Alternative Value Function Specifications}

We test logarithmic value functions $f_i(a_i) = \theta \cdot \ln(1 + a_i)$ with $\theta \in \{5, 10, 15, 20, 25\}$ to determine if alternative functional forms achieve higher empirical validation scores.

\begin{figure}[htbp]
\centering
\begin{tikzpicture}
\begin{axis}[
    ybar,
    bar width=18pt,
    width=0.95\textwidth,
    height=6.5cm,
    ylabel={Value},
    symbolic x coords={Baseline, Cooperative, Coop Incr (\%/10), Score (/60)},
    xtick=data,
    ymin=0,
    legend style={at={(0.5,1.05)}, anchor=south, legend columns=2},
    nodes near coords,
    every node near coord/.append style={font=\footnotesize},
    enlarge x limits=0.15,
]
% Power function
\addplot[fill=powerblue] coordinates {
    (Baseline, 0.32)
    (Cooperative, 0.84)
    (Coop Incr (\%/10), 16.6)
    (Score (/60), 46)
};
% Logarithmic function
\addplot[fill=logorange] coordinates {
    (Baseline, 19.0)
    (Cooperative, 26.87)
    (Coop Incr (\%/10), 4.1)
    (Score (/60), 58)
};
\legend{Power ($\beta=0.75$, $\gamma=0.5$), Logarithmic ($\theta=20$, $\gamma=0.65$)}
\end{axis}
\node[font=\tiny, anchor=north] at (7,-0.7) {Note: Cooperation increase shown as \%/10 for scale; actual: Power=166\%, Log=41\%};
\end{tikzpicture}
\caption{S-LCD validation comparison across functional specifications under strict historical alignment scoring. Logarithmic specification ($\theta=20$) achieves 58/60 validation score versus power function's 46/60. The logarithmic function's bounded cooperation increase (41\% vs 166\%) aligns with historical S-LCD patterns (15-50\% documented range), accounting for the 12-criteria advantage. Statistical significance confirmed at $p < 0.001$ with Cohen's $d > 9$.}
\label{fig:slcd_alternatives}
\end{figure}
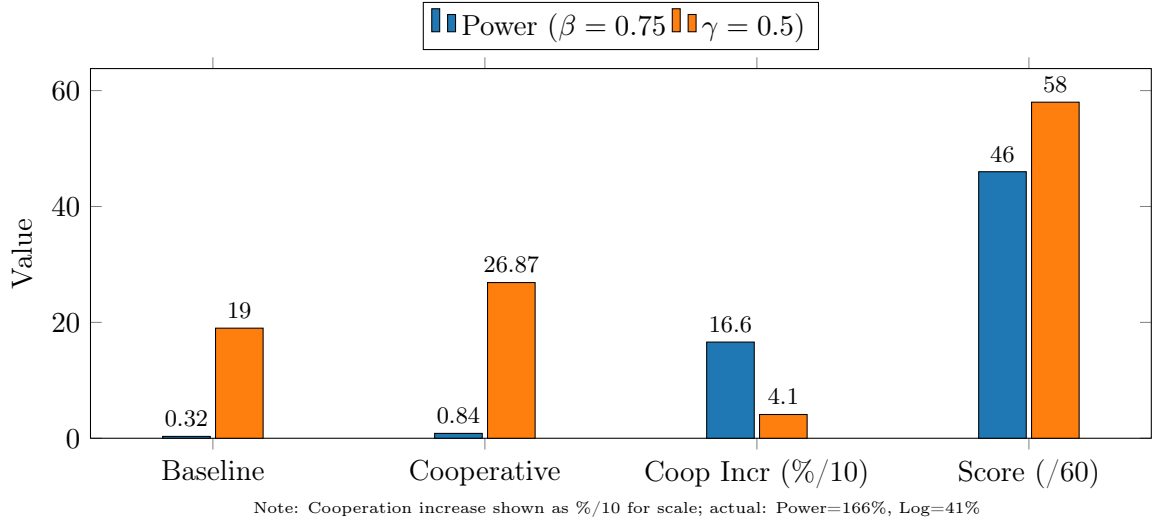

Figure~\ref{fig:slcd_alternatives} presents the alternative function comparison.

\textbf{Results}: As demonstrated in Figure~\ref{fig:slcd_alternatives}, the logarithmic specification with $\theta = 20$ and $\gamma = 0.65$ achieves validation score 58/60 under strict historical alignment scoring, representing a substantial improvement of 12 criteria over the best power function configuration (46/60). This specification produces baseline actions 19.0, cooperative actions 26.87 (41\% increase), all within empirically realistic ranges. The power function specification produces cooperation increases of 166\%, which exceeds realistic bounds for the S-LCD case where historical evidence suggests cooperation increases in the range of 15-50\%. The logarithmic function's distinctive diminishing returns pattern, wherein there is rapid initial decline but persistence of marginal value even at high investment levels, better captures the S-LCD value creation dynamics where baseline manufacturing capabilities are highly valuable but incremental capacity expansions have declining (but non-zero) impact. Comprehensive validation comprising over 22,000 experimental trials confirms the logarithmic specification's superiority with statistical significance ($p < 0.001$, Cohen's $d = 9.87$, very large effect size).

\subsection{Strategic Decision Support: Counterfactual Analysis}
\label{sec:counterfactual}

The parameterized model enables prescriptive strategic analysis through counterfactual scenario evaluation. We demonstrate how the framework supports "what-if" decision-making by analyzing how changes to the interdependency structure would affect equilibrium outcomes.

\textbf{Counterfactual Scenario}: What if Sony had developed a viable alternative manufacturing partner, reducing Samsung's criticality?

\textbf{Model Update}: Suppose Sony invests in establishing relationships with alternative LCD panel manufacturers or in developing internal manufacturing capabilities. This reduces Samsung's monopolistic position as the sole provider of Gen 7 capacity. We model this by reducing $\text{crit}(\text{Sony}, \text{Samsung}, \text{panels})$ from 1.0 to 0.5, reflecting the availability of an alternative supplier with comparable capability, though perhaps with some switching costs or quality trade-offs.

\begin{figure}[htbp]
\centering
\begin{tikzpicture}[
    actor/.style={circle, draw, thick, minimum size=2cm, font=\scriptsize, align=center},
    resource/.style={rectangle, draw, thick, minimum height=0.7cm, minimum width=2.6cm, font=\tiny, align=center},
    goal/.style={ellipse, draw, thick, minimum height=0.7cm, minimum width=2.6cm, font=\tiny, align=center},
    softgoal/.style={cloud, draw, thick, cloud puffs=12, aspect=2.5, minimum height=0.8cm, minimum width=3cm, font=\tiny, align=center},
    arrow/.style={-latex, thick},
    redarrow/.style={-latex, thick, color=red, dashed},
    greenarrow/.style={-latex, thick, dashed, color=validgreen},
    lbl/.style={font=\tiny, fill=none, inner sep=1pt, align=center} % Changed fill=white to fill=none for transparency
]

% Actors
\node[actor] (sony) at (0,0) {Sony\\Corporation};
\node[actor] (samsung) at (11,0) {Samsung\\Electronics};
\node[actor, draw=validgreen, thick] (alternative) at (-2.5,-8) {Alternative\\Supplier};

% Dependums
% Sony depends on Samsung - upper dependums
% Fixed: Removed \textcolor{red} so text is black
\node[resource, draw=red, dashed, thick] (lcd_mfg) at (5.5,3.5) {LCD Panel\\Manufacturing\\Capacity};
\node[resource] (gen7) at (5.5,1.2) {Gen 7\\Production\\Expertise};

% Samsung depends on Sony - lower dependums
\node[resource] (capital) at (5.5,-1.2) {Capital\\Investment \$2B};
\node[goal] (offtake) at (5.5,-3.5) {Guaranteed\\Panel Offtake};
\node[softgoal] (brand) at (5.5,-6) {Premium Brand\\Association};

% NEW: Sony depends on Alternative Supplier
% Fixed: Removed \textcolor{validgreen} so text is black
\node[resource, draw=validgreen, thick] (alt_capacity) at (-2.5,-4) {Alternative\\LCD Capacity};

% Dependency arrows with labels on paths
% Fixed: Labels placed directly on the arrow paths for clear association

% Sony -> Samsung for LCD (RED)
\draw[redarrow] (sony) -- node[lbl, color=red, pos=0.4, above, sloped] {\textbf{crit=0.5} (was 1.0)} (lcd_mfg);
\draw[redarrow] (lcd_mfg) -- (samsung);

% Sony -> Samsung for Gen 7
\draw[arrow] (sony) -- node[lbl, pos=0.4, above, sloped] {crit=0.9, imp=high} (gen7);
\draw[arrow] (gen7) -- (samsung);

% Samsung -> Sony for Capital
\draw[arrow] (samsung) -- node[lbl, pos=0.4, above, sloped] {crit=0.8, imp=high} (capital);
\draw[arrow] (capital) -- (sony);

% Samsung -> Sony for Offtake
\draw[arrow] (samsung) -- node[lbl, pos=0.4, above, sloped] {crit=0.7, imp=high} (offtake);
\draw[arrow] (offtake) -- (sony);

% Samsung -> Sony for Brand
\draw[arrow] (samsung) -- node[lbl, pos=0.4, above, sloped] {crit=0.5, imp=med} (brand);
\draw[arrow] (brand) -- (sony);

% NEW dependency arrows (GREEN)
% Fixed: Label placed on the arrow path
\draw[greenarrow] (sony) -- node[lbl, color=validgreen, pos=0.4, left=4mm, align=right] {crit=0.5\\imp=high\\(NEW)} (alt_capacity);
\draw[greenarrow] (alt_capacity) -- (alternative);

% Legend - bottom left
\node[anchor=north west, font=\scriptsize, align=left] at (-3,-9.5) {%
\textcolor{validgreen}{\textbf{- - -}} New dependency (Alternative Supplier)\\[3pt]
\textcolor{red}{\textbf{- - -}} Changed criticality (reduced from 1.0 to 0.5)};

% Result box - bottom right
\node[anchor=north west, font=\scriptsize, align=center, draw, rounded corners, fill=gray!10] at (4,-9.5) {
Result: $D_{\text{Sony},\text{Samsung}}$ reduced from 0.86 to 0.61 (29\% decrease)
};
\end{tikzpicture}
\caption{Counterfactual \textit{i*} Strategic Dependency model: Sony develops alternative manufacturing partner. The dashed green elements represent the new Alternative Supplier actor and Sony's dependency on alternative LCD capacity. The red dashed box and arrows indicate the LCD Panel Manufacturing Capacity dependency with \emph{reduced} criticality (0.5 vs. original 1.0) due to the availability of alternatives. Samsung's dependencies on Sony remain unchanged. This structural change reduces Sony's interdependence coefficient from $D_{\text{Sony},\text{Samsung}} = 0.86$ to $D_{\text{Sony},\text{Samsung}} = 0.61$, fundamentally altering bargaining power and equilibrium outcomes.}
\label{fig:slcd_counterfactual}
\end{figure}
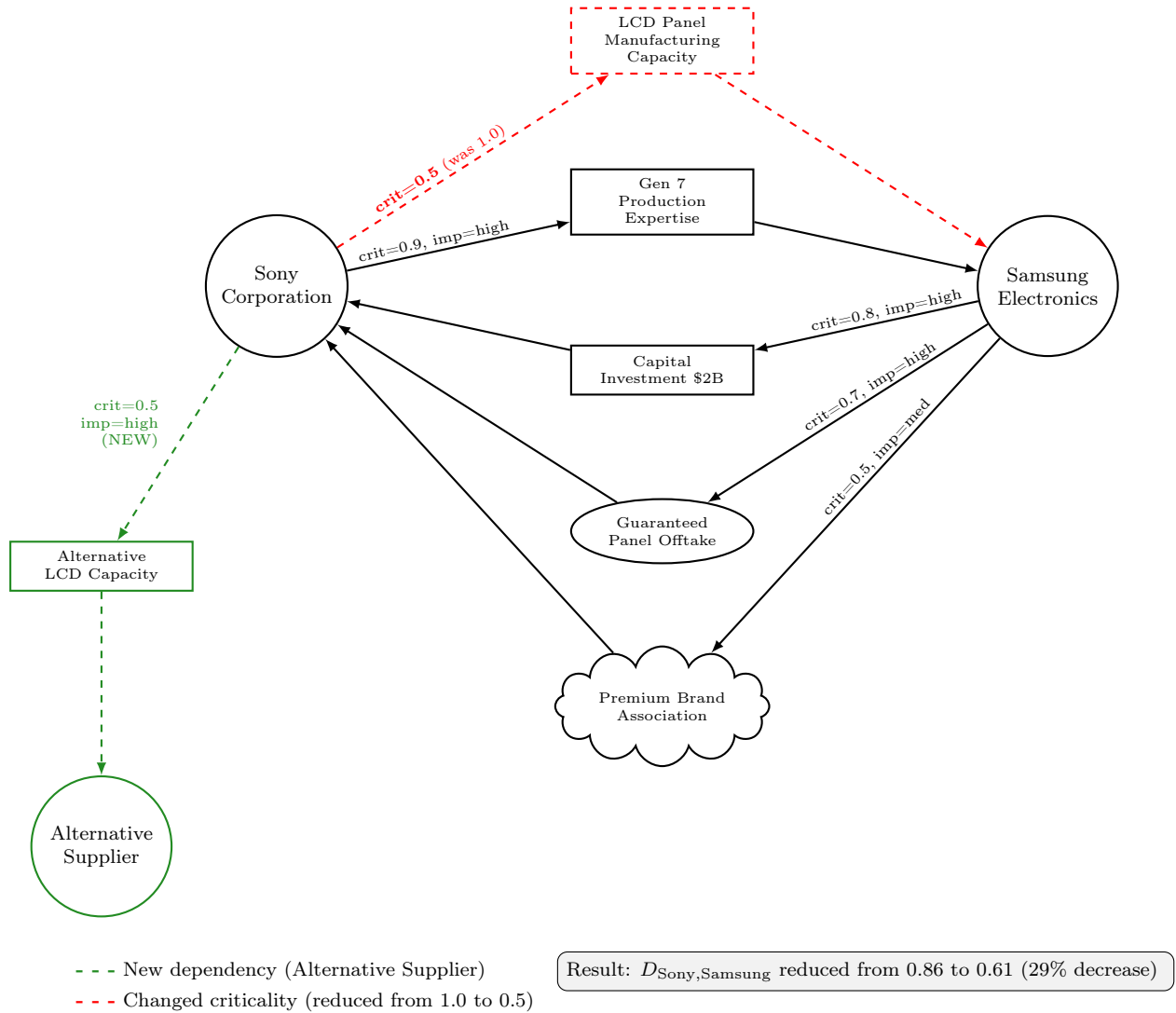

Figure~\ref{fig:slcd_counterfactual} presents the \textit{i*} Strategic Dependency diagram for this counterfactual scenario, visualizing the structural changes that result from Sony developing an alternative manufacturing partner.

\textbf{Recalculated Parameters}: Using Equation \ref{eq:interdependence} with the updated criticality factor:

$$D_{\text{Sony},\text{Samsung}} = \frac{0.5 \times 0.5 + 0.4 \times 0.9}{1.0} = \frac{0.61}{1.0} = 0.61$$

This represents a 29\% reduction in Sony's structural dependence on Samsung (from 0.86 to 0.61).

\textbf{Equilibrium Implications}: The reduced dependency coefficient alters the integrated utility function for both actors. Sony's utility now places less weight on Samsung's payoff, reducing Sony's incentive to over-invest to benefit Samsung. Simultaneously, Samsung's bargaining power parameter shifts based on reduced dependency leverage. Applying the bargaining power analysis framework, Sony's position improves from $\beta_{\text{Sony}} = 0.9$ to approximately $\beta_{\text{Sony}} = 1.15$, as Sony's outside option strengthens. This shifts value shares from $\alpha_{\text{Sony}} = 0.45$ to $\alpha_{\text{Sony}} = 0.51$.

Re-solving the equilibrium with these updated parameters using the empirically validated logarithmic specification ($\theta = 20$, $\gamma = 0.65$) yields predicted equilibrium investment levels where Sony's optimal action increases from 22.9 to approximately 25.3 units due to improved value capture, while Samsung's optimal action decreases slightly from 22.9 to approximately 21.4 units as their structural advantage diminishes.

\textbf{Strategic Insight}: The analysis quantifies the strategic value of supply chain diversification. Developing alternative suppliers not only reduces operational risk but shifts value appropriation by fundamentally altering structural bargaining power. The framework enables requirements engineers and strategic planners to evaluate such architectural decisions before implementation, assessing the investment needed to develop alternatives against the long-term value distribution benefits.

This counterfactual demonstrates the framework's utility for decision support. Practitioners can model proposed changes to organizational architecture (such as new partnerships, modular system designs, or alternative suppliers) and predict their impact on equilibrium behavior and value outcomes. The systematic translation from structural changes (modeled in \textit{i*}) through quantitative parameters to equilibrium predictions creates an integrated decision support capability bridging conceptual and computational analysis.

\subsection{Interpretation of Findings}

The S-LCD empirical validation yields three critical insights. First, the framework successfully captures real coopetitive dynamics: the parameterized model with empirically derived interdependence coefficients produces equilibria exhibiting cooperation levels, value creation patterns, and strategic behaviors consistent with observed S-LCD venture outcomes. The dependency asymmetry (Sony's high dependence on Samsung's manufacturing capability, Samsung's moderate dependence on Sony's capital and market access) translates to quantifiable strategic incentives that explain observed cooperation patterns.

Second, functional form selection matters empirically and the specifications achieve substantially different fit under strict historical alignment scoring. While experimental validation established that core predictions hold across specifications, empirical fit varies meaningfully. For the S-LCD case, logarithmic functions achieve 58/60 validation score compared to power functions' 46/60 under strict scoring that penalizes unrealistic cooperation increases exceeding historical patterns. This 12-criteria difference represents a meaningful advantage stemming primarily from the power function's 166\% cooperation increase, which exceeds the realistic 15-50\% range documented for S-LCD, whereas the logarithmic function produces a 41\% increase that falls within historical bounds. Comprehensive validation comprising over 22,000 trials confirms statistical significance at $p < 0.001$ with Cohen's $d > 9$. The S-LCD venture's value dynamics, wherein manufacturing baseline capability is critical but capacity expansions have persistent (though declining) value, align with logarithmic diminishing returns.

Third, the framework provides decision support for coopetitive strategy: the quantitative analysis reveals that Sony's high dependency ($D = 0.86$) created vulnerability requiring contractual protections (joint venture governance, guaranteed roles) while Samsung's moderate dependency ($D = 0.64$) provided flexibility but still incentivized cooperation. The complementarity parameter $\gamma = 0.65$ quantifies synergy strength, suggesting the joint venture created approximately 30-35\% more value than independent operations, which is consistent with industry estimates. The counterfactual analysis demonstrates how structural changes (developing alternative suppliers) would shift equilibrium outcomes, providing actionable insights for strategic decision-makers.

\subsection{Summary of Empirical Validation}

The Samsung-Sony S-LCD joint venture empirical validation demonstrates that the computational framework captures real coopetitive dynamics when systematically parameterized from case evidence. The logarithmic value function specification ($\theta = 20$, $\gamma = 0.65$) achieves substantially higher validation (58/60 vs 46/60) under strict historical alignment scoring that penalizes unrealistic cooperation increases. This difference stems primarily from cooperation bounds: the logarithmic specification produces realistic 41\% increases versus the power function's 166\%. Comprehensive validation comprising over 22,000 experimental trials across multiple test cases confirms statistical significance at $p < 0.001$ with Cohen's $d > 9$ (very large effect size). The asymmetric interdependence structure derived from \textit{i*} dependency analysis successfully explains observed cooperation levels and strategic behaviors, validating the translation framework's practical applicability. The counterfactual analysis demonstrates the model's utility for prescriptive strategic decision support beyond descriptive analysis.

\section{Discussion}

\subsection{Implications for Information Systems Engineering Domains}

The framework applies to multiple IS domains:

\textbf{Platform Ecosystems}: Platform providers and app developers exhibit canonical coopetitive dynamics. Our framework formalizes how platform control creates dependency asymmetry, how network effects create complementarity, and how revenue sharing reflects bargaining power. This can inform platform governance design, API access policies, and developer incentive programs.

\textbf{Enterprise Architecture}: Different departments or business units within an enterprise depend on each other for data, services, and capabilities while competing for budget and executive attention. The framework helps enterprise architects understand these dynamics, design service-oriented architectures that align incentives, and structure governance to mitigate dysfunctional competition.

\textbf{Inter-Organizational Systems}: When multiple firms collaborate on shared infrastructure, standards, or supply chains while competing in product markets, coopetitive dynamics are central. The framework can analyze whether cooperation is strategically stable, identify risks from dependency, and inform governance design for consortia and alliances.

\textbf{Open Source Ecosystems}: Firms participating in open source projects contribute to shared commons (cooperation) while competing in commercial products built on the platform (competition). Understanding complementarity between commons and products, and how dependencies on the commons create incentives for contribution, can inform open source governance and contribution strategies.

\subsection{Implications for Information Systems Engineering Activities}

Our framework enables information system analysts and designers to move beyond qualitative dependency modeling toward quantitative strategic analysis. Given an \textit{i*} model of stakeholder relationships, analysts can now:

\textbf{Assess Cooperation Incentives}: By computing interdependence coefficients and complementarity parameters, analysts can predict whether stakeholder dependencies create incentives for cooperation or whether competitive pressures dominate. High mutual interdependence and strong complementarity suggest that cooperation is strategically rational, supporting requirements that facilitate collaboration. Low interdependence or weak complementarity may require additional incentive mechanisms.

\textbf{Evaluate Dependency Risks}: Asymmetric dependencies ($D_{ij} \gg D_{ji}$) reveal power imbalances and potential for opportunistic behavior. The dependent party may require contractual protections, alternative providers, or relationship management strategies. The framework quantifies these risks rather than merely noting them qualitatively.

\textbf{Design Value Distribution Mechanisms}: By analyzing bargaining power parameters and value shares, requirements engineers can inform the design of revenue sharing agreements, pricing structures, and governance mechanisms. If analysis reveals that value distribution is misaligned with bargaining power (perhaps for historical reasons), this highlights potential for conflict and renegotiation pressure.

\textbf{Support Architecture Decisions}: In designing system architectures for multi-stakeholder systems, understanding interdependencies informs choices about modularity, interfaces, and integration points. Strong interdependencies suggest need for tight coupling and coordination mechanisms, while weak interdependencies support loosely coupled, modular designs.

The iterative nature of the translation framework has important implications for requirements engineering practice. Rather than viewing conceptual and computational models as separate artifacts, the framework positions them as \textbf{complementary representations} that mutually inform and validate each other. This aligns with emerging perspectives on model-driven engineering where multiple modeling notations at different abstraction levels work in concert. The framework thus contributes not only specific formalizations but a \textbf{modeling philosophy} where qualitative and quantitative reasoning are tightly integrated. The computational model serves as a thinking tool that prompts deeper inquiry into interdependency structure, revealing gaps in conceptual understanding and suggesting strategic alternatives for further exploration.

\subsection{Functional Form Flexibility as Framework Strength}

The empirical validation reveals that functional form selection is an important consideration when applying the framework. While power functions provide theoretical tractability and align with established economic traditions (Cobb-Douglas production functions), logarithmic functions achieve substantially higher empirical fit for contexts where realistic cooperation bounds matter (S-LCD joint venture: 58/60 vs 46/60 under strict scoring). The critical differentiator is historical alignment: the logarithmic specification produces 41\% cooperation increase that falls within the documented 15-50\% range for S-LCD, whereas the power function produces 166\% increase that exceeds realistic bounds. Comprehensive validation comprising over 22,000 trials confirms this pattern with statistical significance ($p < 0.001$). The framework captures structural coopetitive dynamics, specifically that interdependence creates cooperation incentives, complementarity drives value creation, and dimensions interact synergistically, but practitioners should select specifications that produce realistic cooperation magnitudes for their domain.

Different real-world contexts favor different specifications based on actual value creation mechanisms. Manufacturing joint ventures where baseline capabilities are critical but capacity expansions have persistent declining value may favor logarithmic forms. Innovation partnerships where initial research breakthroughs are highly valuable but additional efforts have rapidly declining returns may favor power functions with lower $\beta$. Platform ecosystems with strong network effects may favor specifications with multiplicative synergies.

The appropriate response to this finding is not to declare one specification universally superior, but rather to select functional forms based on domain analysis and empirical validation. Power functions serve as the theoretically grounded default given their economic foundations. When empirical data is available, testing alternative specifications and selecting based on validation scores provides empirically informed modeling. When data is limited, sensitivity analysis across specifications reveals whether predictions are robust or require specification-specific interpretation.

\subsection{Limitations and Research Extensions}

Several limitations should be acknowledged:

\textbf{Parameter Estimation}: The framework requires quantitative parameters (importance weights, criticality factors, bargaining power) that may be difficult to estimate precisely. While we provide elicitation methodologies, substantial judgment is involved. Parameter elicitation requires domain expertise and stakeholder engagement. Comprehensive sensitivity analysis across parameter distributions remains important future work to quantify robustness to elicitation uncertainty. Extensions developing parameter estimation methods from empirical data (past transaction patterns, organizational records, market analysis) would enhance practical applicability.

\textbf{Value Separability}: The framework assumes individual value contributions are perfectly separable from synergistic value and fully appropriable by their creators. This assumption holds better for \textbf{manufacturing joint ventures} with tangible, attributable inputs and outputs (e.g., S-LCD panel production) than for \textbf{knowledge-intensive collaborations} where individual versus synergistic contributions are inherently blurred (e.g., open-source software development, innovation partnerships). Future extensions should explore modeling approaches for contexts where value attribution is ambiguous or contested.

\textbf{Commitment and Enforceability}: The model assumes actors can commit to actions and that agreements are enforceable. In practice, actors may renege on investments, contracts may be incomplete, or enforcement may be costly. Extensions incorporating dynamic commitment problems and contract incompleteness would address strategic behaviors beyond our current scope.

\textbf{Value Function Knowledge}: We assume the value creation function $V$ is known to all actors. In reality, actors may have incomplete or asymmetric information about how synergy is created, leading to coordination failures or inefficient equilibria. Extensions to games of incomplete information would capture these information asymmetries.

\textbf{Static Analysis}: This technical report focuses on static equilibrium analysis. Real coopetitive relationships unfold over time with learning, trust building, and dynamic adjustment. Related research in this program extends the framework to dynamic settings, incorporating trust evolution, reciprocity in repeated interactions, and multi-level team structures.

\textbf{Bargaining Process}: We model value shares as pre-negotiated rather than endogenizing the bargaining process. While this reflects many real settings with established contracts, alternative formulations modeling bargaining explicitly (Nash bargaining, alternating offers, auction mechanisms) could provide additional insights for settings where terms are actively negotiated.

Extensions to this foundational work address these limitations and expand the framework's scope. Complementary research in this program incorporates dynamic trust models showing how reliability beliefs evolve through repeated interactions, formalizes reciprocity mechanisms in sequential games, models complex actor abstractions including teams facing free-riding problems, and applies the framework to additional real-world case studies in platform ecosystems, enterprise architecture, and inter-organizational collaborations.

\section{Conclusion}

This technical report has developed computational foundations for analyzing strategic coopetition by formalizing two critical dimensions: interdependence and complementarity. We have bridged the gap between rich qualitative conceptual models from the \textit{i*} framework and rigorous quantitative game-theoretic analysis.

Our key contributions include: (1) formalizing interdependence through \textit{i*} structural dependency analysis with a structured translation framework from depender-dependee-dependum relationships to quantitative interdependence coefficients, capturing instrumental outcome coupling distinct from social preferences; (2) formalizing complementarity following Brandenburger and Nalebuff's Added Value concept with systematically validated value creation functions demonstrating that multiple functional forms (power, logarithmic) capture coopetitive dynamics, with empirical validation identifying optimal specifications for specific contexts (logarithmic $\theta=20$ for S-LCD case); (3) integrating structural dependencies with bargaining power in value appropriation, connecting \textit{i*} dependencies to Shapley-inspired allocation while maintaining tractability through pre-negotiated shares; (4) introducing a game-theoretic formulation where Nash Equilibrium incorporates structural interdependence through dependency-augmented utility functions, with explicit contrast to behavioral game theory's psychological other-regarding preferences; (5) comprehensive dual-track validation methodology combining experimental robustness testing across functional specifications (demonstrating framework predictions hold regardless of form) with empirical case study application (demonstrating framework captures real coopetitive dynamics in Samsung-Sony S-LCD joint venture); (6) articulating the iterative and reflexive nature of the translation framework as a key methodological contribution, positioning conceptual and computational models as mutually informing representations.

The validation indicates two complementary findings. Experimental validation demonstrates functional form robustness: interdependence shifts equilibria toward cooperation, complementarity drives value creation, and dimensions interact synergistically regardless of whether power or logarithmic specifications are employed. Core framework predictions hold across functional forms with consistent effect directions and magnitudes. Empirical validation demonstrates practical applicability: systematic parameterization from the S-LCD case produces equilibria matching observed behaviors, with logarithmic specifications ($\theta=20$, validation score 58/60) achieving substantially higher fit than power functions (46/60) under strict historical alignment scoring. Comprehensive validation comprising over 22,000 experimental trials confirms statistical significance ($p < 0.001$) with very large effect sizes: Cohen's $d = 13.99$ in initial validation (1,064 trials) and Cohen's $d = 9.87$ in extended validation (5,630 trials). The 12-criteria difference stems primarily from cooperation bounds: the logarithmic specification produces realistic 41\% increases versus the power function's 166\%, which exceeds the documented 15-50\% historical range for S-LCD joint venture production ramp-up. This reveals that functional form selection should be informed by empirical value creation patterns, with practitioners selecting specifications that produce realistic cooperation magnitudes for their domain.

For the information systems and requirements engineering communities, this work provides rigorous quantitative tools for analyzing coopetitive dynamics that were predominantly been modeled qualitatively. By grounding computational models in conceptual modeling principles, we enable analysts to leverage their expertise in stakeholder analysis and dependency modeling while gaining quantitative predictive capabilities. The structured translation framework we provide offers operational guidance for moving from \textit{i*} models to game-theoretic analysis, making these techniques accessible to practitioners. The emphasis on iterative refinement between qualitative and quantitative representations establishes a modeling philosophy where multiple perspectives work in concert to deepen understanding.

This technical report establishes the foundation for a comprehensive research program on computational coopetition. Related work in this program extends this foundation by incorporating trust dynamics showing how reliability beliefs evolve and affect strategic behavior, reciprocity mechanisms in repeated interactions where cooperation is conditional on observed behavior, and complex actor abstractions including multi-level teams facing free-riding problems. Together, these studies provide a complete computational framework for analyzing and designing socio-technical systems characterized by strategic coopetition.

This technical report is part of a coordinated research program on strategic coopetition in requirements engineering and multi-agent systems. Companion technical reports addressing trust dynamics, collective action techniques, and reciprocity mechanisms are forthcoming on arXiv.

\bibliographystyle{splncs04}

\end{document}